\documentclass[aps,prl,reprint,10pt,twocolumn,nofootinbib,longbibliography,superscriptaddress]{revtex4-1}
\usepackage{amsmath}
\usepackage{amsfonts}
\usepackage{amssymb}
\usepackage{graphicx}
\usepackage{epstopdf}
\usepackage{color}
\usepackage{appendix}
\usepackage{csvsimple}
\usepackage{chngpage}
\usepackage{subcaption}
\usepackage{mathrsfs}
\usepackage{booktabs}
\usepackage{float}
\usepackage[cm]{fullpage}
\usepackage[colorlinks,linkcolor=blue,citecolor=blue,urlcolor=blue]{hyperref}
\usepackage[table,xcdraw]{xcolor}
\usepackage[utf8]{inputenc}
\newcommand{\mpl}{m_\text{P}}

\newcommand{\pip}{\phi_{\mathrm{IP}}}
\newcommand{\nn}{\nonumber}

\begin{document}

\title{\boldmath{Instant Preheating in Quintessential Inflation with $\alpha-$Attractors}}
\date{\today}
\author{Konstantinos Dimopoulos}
\affiliation{Consortium for Fundamental Physics, Physics Department,Lancaster University, Lancaster LA1 4YB, United Kingdom}
\author{Leonora Donaldson Wood}
\affiliation{Consortium for Fundamental Physics, Physics Department,Lancaster University, Lancaster LA1 4YB, United Kingdom}
\author{Charlotte Owen}
\affiliation{Consortium for Fundamental Physics, Physics Department,Lancaster University, Lancaster LA1 4YB, United Kingdom}

\begin{abstract}
We investigate a compelling model of quintessential inflation in the context of
$\alpha$-attractors, which naturally result in a scalar potential featuring
two flat regions; the inflationary plateau and the quintessential tail.
The ``asymptotic freedom'' of $\alpha$-attractors, near the kinetic
poles, suppresses radiative corrections and interactions, which would otherwise
threaten to lift the flatness of the quintessential tail and cause a 5th-force 
problem respectively. Since this is a non-oscillatory inflation model, we reheat
the Universe through instant preheating. The parameter space is constrained by
both inflation and dark energy requirements. We find an excellent correlation
between the inflationary observables and model predictions, in agreement with 
the $\alpha$-attractors set-up. We also obtain successful quintessence for 
natural values of the parameters. Our model predicts potentially sizeable tensor
perturbations (at the level of 1\%) and a slightly varying equation of state for
dark energy, to be probed in the near future.
\end{abstract}

\maketitle

\nopagebreak
\section{Introduction}
One of the greatest discoveries in cosmology was that the Universe is currently 
undergoing accelerated expansion \cite{Riess:1998cb,Perlmutter:1998np}. To 
account for this, Einstein's General Relativity demands that the dominant 
component of the Universe content violates the strong energy condition. Assuming
that it is a barotropic fluid, its pressure must be negative enough
\mbox{$p<-\frac13\rho$}. Such a mysterious substance is dubbed `dark energy'.

By far, the simplest choice of dark energy is vacuum density, due to
a non-zero cosmological constant, for which \mbox{$p=-\rho$}. The main 
problem with this idea is that the required value of the cosmological constant
is staggeringly small such that the vacuum density is about $10^{120}$ times
smaller than the Planck density%
, which corresponds to the cutoff scale of the theory.
This has been called ``the worst fine-tuning in Physics''.%
\footnote{by Laurence Krauss.} To overcome this, but at the expense of 
introducing new Physics, there have been alternative proposals put forward.

A substance which can exhibit pressure negative enough is a potentially dominated homogeneous scalar field. Thus, it is possible for a dynamical scalar field to 
drive the accelerated expansion of the Universe, therefore being a type of dark 
energy. The idea has long been in use when modelling cosmic inflation in the 
early Universe (inflationary paradigm). Mirroring the mechanism used to explain 
inflation, this idea can also be applied to address the current accelerated 
expansion. A scalar field responsible for late inflation is called quintessence;
the fifth element after baryons, CDM, photons and neutrinos \cite{Peebles:2002gy,Ratra:1987rm,Caldwell:1997ii}.

Since they are both based on the same idea, it is natural to attempt to unify
cosmic inflation with quintessence. Indeed, the mechanism in which a scalar 
field is both driving primordial inflation and causing the current accelerated 
expansion, is called quintessential inflation \cite{Peebles:1998qn}. 
Quintessential inflation is economical in that it models both inflation and 
quintessence in a common theoretical framework and employs a single degree of 
freedom. It also features some practical advantages, for example the initial
conditions of quintessence are determined by the inflationary attractor. As such
the infamous coincidence problem (which corresponds to late inflation occurring 
at present) is reduced to a constraint on the model parameters and not on 
initial conditions.

Quintessential inflation models require the inflaton potential energy density to
survive until the present day to act as dark energy. Amongst other things, this 
necessitates a reheating mechanism alternative to the standard assumption, in 
which, after the end of inflation, the inflaton field decays into the thermal 
bath of the hot big bang. If inflaton decay is not considered, then reheating
must occur by other means.
Different reheating alternatives to inflaton decay include
instant preheating \cite{Felder:1998vq,Campos:2002yk}, curvaton reheating 
\cite{Feng:2002nb,BuenoSanchez:2007jxm,Matsuda:2007ax} and gravitational 
reheating \cite{Ford:1986sy,Chun:2009yu}, amongst others \cite{Copeland:2006wr}, 
which may or may not happen exclusively. For example, gravitational reheating 
is \textit{always} present, but because it is a very inefficient mechanism it 
is overwhelmed if another reheating mechanism is present.
%
The outcome of any of these reheating mechanisms must complete and lead to 
radiation domination well before the time of Big Bang Nucleosynthesis (BBN).
The temperature of the Universe when radiation domination takes over is
called the reheating temperature $T_{\mathrm{reh}}$. Thus we need 
\mbox{$T_{\rm reh}\gg 1\,$MeV} to avoid disturbing BBN.

In quintessential inflation models, if reheating is not prompt, a period of 
kination exists, where the kinetic density of the inflaton is the 
dominant energy density in the Universe. During this period, the non-decaying 
mode for gravitational waves is not suppressed and this produces a spike in the 
gravitational wave spectrum at high frequencies. The energy density of these 
gravitational waves may be large enough to disturb the BBN process. As such it
is constrained and provides an upper limit on the duration of kination. This, in 
turn, is a lower bound on the reheating temperature. 

The scalar potential in quintessential inflation typically features two flat 
regions, which when traversed by the scalar field, result in accelerated 
expansion, provided that the scalar field dominates the Universe. These are 
called the inflationary plateau and the quintessential tail and can lead to
the primordial and current accelerated expansion respectively. Thus, the
required potential is of runaway type, with the global minimum displaced at 
infinity, where the vacuum density is zero. The form of the quintessential 
potential is non trivial, especially since the two plateaus differ by more than
a factor of $10^{100}$ in energy density. Consequently, one needs to use a 
theoretical framework that is valid at both these extreme energy scales.

A compelling way to naturally generate a scalar potential with the desired
two plateaus is the idea of $\alpha$-attractors, which is heavily used in 
inflationary model building \cite{Kallosh:2013yoa,Kallosh:2013tua,Ferrara:2013rsa,Ferrara:2013kca,Kallosh:2014rga,Kallosh:2015lwa,Roest:2015qya,Linde:2015uga,Scalisi:2015qga,Carrasco:2015pla,Galante:2014ifa,Carrasco:2015rva,Terada:2016nqg,Kumar:2015mfa,Ueno:2016dim,Eshaghi:2016kne,Artymowski:2016pjz,Kallosh:2016sej,Ferrara:2016fwe,Odintsov:2016vzz,DiMarco:2017sqo,Alho:2017opd,Dimopoulos:2016yep}. Recently, we have presented 
a new quintessential inflation model along these lines 
\cite{Dimopoulos:2017zvq}. Our model is in excellent agreement for inflationary 
observables with the CMB observations \cite{Ade:2015lrj}. In our paper in 
Ref.~\cite{Dimopoulos:2017zvq} we utilised the mechanism of gravitational 
reheating to reheat the Universe after inflation, in alignment with the economy 
of the quintessential inflation idea. 
However, gravitational reheating is notoriously inefficient, with a very low 
reheating temperature of approximately $T_{\mathrm{reh}}\sim 10^{4}\,\mathrm{GeV}$.
As a result, the spike of gravitational waves due to kination is large enough 
to challenge the BBN process. This is the price to pay for economy. In this 
paper we generalise our approach and employ the instant preheating mechanism 
to reheat the Universe. Thus, we envisage a coupling between our scalar 
field with some other degree of freedom such that, after the end of inflation,
the rapid variation of the inflaton's expectation value leads to 
non-perturbative particle production, which generates the radiation bath of the
hot big bang. The process is modulated by the coupling constant $g$. If $g$ is 
small enough, instant preheating becomes comparable to gravitational reheating.
Thus, in general we consider \mbox{$T_{\rm reh}>10^4\,$GeV}.

The setup of $\alpha$-attractors also has another beneficial 
consequence, apart from generating the potential plateaus. It has to do with
the suppression of radiative corrections near the poles (so along the plateaus),
which otherwise threaten to lift the flatness of the quintessential tail, as 
well as the suppression of interactions, which would otherwise generate a 5th-force problem for
quintessence. Both these problems plague models of quintessence and 
quintessential inflation alike. In Ref.~\cite{Dimopoulos:2017zvq}, we had not 
fully realised the beneficial effect of the $\alpha$-attractors setup in this 
respect. Thus, we aimed to avoid excessive interactions by keeping the 
non-canonical field sub-Planckian (the interactions are Planck-suppressed). 
However, as demonstrated in Refs.~\cite{Linde:2017pwt,Kallosh:2016gqp}, the suppression of loop 
corrections and interactions along the plateaus (near the poles) is such that 
even a super-Planckian excursion of the non-canonical inflaton is admissible. 
We investigate this issue in detail, but conservatively choose to avoid
super-Planckian values for our non-canonical inflation field in our treatment.

We start with an overview of the model before noting how a change of reheating 
mechanism affects the number of remaining inflationary e-folds since observable 
scales left the horizon during primordial inflation. We calculate the 
inflationary observables before investigating how the quintessence requirements 
determine the parameter space in a model with instant preheating. 


We use natural units, where $c=\hbar=1$ and Newton's gravitational 
constant is \mbox{$8\pi G=\mpl^{-2}$}, with \mbox{$\mpl
=2.43\times 10^{18}\,$GeV} being the reduced Planck mass.

	\section{The Model}

\subsection{The Scalar Potential}

We consider the following model:
\begin{equation}\label{L0}
\mathcal{L} = \mathcal{L}_{\rm kin} + \mathcal{L}_V + \Lambda\,,
\end{equation}
where $\mathcal{L}_{\rm kin}$ is the kinetic Lagrangian density, 
$\mathcal{L}_V$ is the potential Lagrangian density and $\Lambda$ is a 
cosmological constant. 

The kinetic Lagrangian density is
\begin{equation}\label{Lkin}
\mathcal{L}_{\rm kin} = \frac{\frac{1}{2}(\partial\phi)^2}
{(1 - \frac{\phi^2}{6\alpha\mpl^2})^2}\,,
\end{equation}
where $\alpha>0$ is a parameter. This is the standard, non-canonical form in the
context of $\alpha$-attractors \cite{Kallosh:2013yoa,Kallosh:2013tua,Ferrara:2013rsa,Ferrara:2013kca}. It can be realised in supergravity 
theories, when the K\"{a}hler manifold is not trivial, such that 
$\mathcal{L}_{\rm kin}$ features poles, characterised by the $\alpha$ parameter.

For the potential Lagrangian density, we consider a simple exponential function
(possibly due to gaugino condensation \cite{Gorlich:2004qm,Haack:2006cy,Lalak:2005hr}). Thus, we have
\begin{equation}\label{LV0}
-\mathcal{L}_V = V(\phi)=V_0e^{-\kappa\phi/\mpl}\,.
\end{equation}
where $\kappa$ is a parameter (without loss of generality, we consider 
\mbox{$\kappa>0$}) and $V_0$ is a constant density scale.

In an effort to minimise its potential density, the expectation value of the 
field $\phi$ grows in time. However, it cannot cross the poles at 
$\pm\sqrt{6\alpha}\,\mpl$ \cite{Kallosh:2013yoa,Kallosh:2013tua,Ferrara:2013rsa,Ferrara:2013kca}. Thus, starting in-between the poles,
we expect that it finally approaches the value 
\mbox{$\phi\rightarrow+\sqrt{6\alpha}\,\mpl$}, which corresponds to non-zero
potential density \mbox{$V(\sqrt{6\alpha}\,\mpl)=V_0e^{-\kappa\sqrt{6\alpha}}$}.

We assume that, due to an {\em unknown} symmetry, the vacuum density is zero.
This was the standard assumption before the discovery of dark energy. If a 
non-zero vacuum density is assumed then we have the usual $\Lambda$CDM 
cosmology. For motivating quintessence as the explanation of the dark energy 
observations, the vacuum density has to be zero. 
This fixes the cosmological constant in our model to the value
\begin{equation}\label{Lambda}
\Lambda = V(\sqrt{6\alpha}\,\mpl)= V_0e^{-\kappa\sqrt{6\alpha}}.
\end{equation}
Defining $n \equiv \kappa\sqrt{6\alpha}$ and incorporating $\Lambda$,
the scalar potential can now be expressed as 
\begin{equation}\label{eq:Scalar_pot}
V(\phi) = V_0e^{-n}
\left[e^{n\left(1-\frac{\phi}{\sqrt{6\alpha}\mpl}\right)}-1\right] 
\,.
\end{equation} 

To assist our intuition, it is useful to consider a canonically normalised
inflaton field $\varphi$. The form of the kinetic Lagrangian density in
Eq.~\eqref{Lkin} suggests the field redefinition is obtained when 
\mbox{$\frac{\partial\phi}{\partial\varphi}=1-\frac{\phi^2}{6\alpha\mpl^2}$}, 
which gives
\begin{equation}\label{eq:canonicalVariables}
\phi=\sqrt{6\alpha}\,\mpl\;\tanh\left(\frac{\varphi}{\sqrt{6\alpha}\mpl}\right)	
\,.
\end{equation}	
Then, the scalar potential, in terms of the canonical scalar field becomes
\begin{equation}\label{V}
V(\varphi) = e^{-2n}M^4 
\Big\{\exp\Big[n\Big(1-\tanh\frac{\varphi}{\sqrt{6\alpha}m_P}\Big)\Big]-1\Big\}	
\quad\,,
\end{equation}
where we have defined \mbox{$M^4 \equiv e^nV_0$}, which stands for the 
inflation energy scale. 
Note, also, that \mbox{$\Lambda=e^{-2n}M^4$}.

Whereas the range of the non-canonical inflaton field $\phi$ is bounded by 
the poles in ${\cal L}_{\rm kin}$: 
\mbox{$-\sqrt{6\alpha}<\phi/\mpl<\sqrt{6\alpha}$}, the range of the canonical
inflaton field $\varphi$ is unbounded: \mbox{$-\infty<\varphi<+\infty$}. This 
is because the poles are transposed to infinity when we switch from $\phi$ to
$\varphi$. In effect, the scalar potential $V(\varphi)$ becomes ``stretched'' 
as $\phi$ approaches the poles \cite{Kallosh:2013yoa,Kallosh:2013tua,Ferrara:2013rsa,Ferrara:2013kca}. Therefore, the potential 
$V(\varphi)$ features two plateaus experienced by the field, one at early and 
one at late times. 

At early times 
\mbox{($\varphi\rightarrow-\infty, \phi\rightarrow-\sqrt{6\alpha}\,\mpl$)}, 
the potential in Eq.~\eqref{V} can be simplified to
\begin{equation}\label{eq:V_Infl}
V(\varphi)\simeq M^4\mathrm{exp}
\Big(-2ne^{\frac{2\varphi}{\sqrt{6\alpha}\mpl}}\Big)\,,
\end{equation}
which gives rise to the inflationary plateau. In the opposite limit, 
towards late times 
\mbox{($\varphi\rightarrow+\infty,\phi\rightarrow+\sqrt{6\alpha}\,\mpl$)}, the 
potential in Eq.~\eqref{V} becomes
\begin{equation}\label{eq:V_quint}
V = 2ne^{-2n}M^4\,\mathrm{exp}(-2\varphi/\sqrt{6\alpha}\,\mpl) \,.
\end{equation}
This corresponds to the quintessential tail. It is evident that the potential 
density asymptotes to zero as \mbox{$\varphi\rightarrow+\infty$}.

The evolution of the quintessential inflaton field goes as follows. The field 
slow-rolls along the early-time plateau, obeying the slow-roll constraints and 
inflating the Universe. Inflation ends when the 
potential becomes steep and curved. Afterwards, the inflaton field falls down 
the steep slope of the potential. A period of kination ensues, when the Universe
is dominated by the kinetic density of the scalar field. Kination ends when the 
Universe is reheated and radiation takes over. As such, the duration of kination
is inversely proportional to the reheating temperature $T_{\mathrm{reh}}$, which 
defines the moment when radiation domination begins and reheating completes. The
field continues to roll until it runs out of kinetic energy and freezes at a 
particular value $\varphi_F$. It remains dormant at $\varphi_F$ until late 
times, when it becomes quintessence and its residual potential density drives 
the Universe expansion into acceleration again.

Here we should highlight the importance of the parameter 
\mbox{$n=\kappa\sqrt{6\alpha}$}. The value of $n$ modulates both the steepness 
of the potential and the inflaton value where the potential drops from the 
early-time to the late-time plateau. As such, this controls $\varphi_F$, the 
value the field freezes at, when it runs out of kinetic energy after reheating. 

\subsection{\boldmath The Range of $\alpha$}

At late times, there are two attractor solutions to the Klein-Gordon equation
depending on whether the quintessence field is eventually dominant or not over 
the background matter. It has been shown that, when the background density 
becomes comparable to the field's residual potential density $V(\varphi_F)$,
the field unfreezes and briefly oscillates about the attractor before settling 
on the attractor solution \cite{Copeland:1997et}. The question of which 
attractor solution the field eventually follows is controlled by 
the value of $\alpha$, which determines the slope of the quintessential tail.

The latest Planck observations suggest that the density parameter of dark energy
is \mbox{$\Omega_\Lambda=1-\Omega_{\mathrm{K}}-\Omega_{m}$}, where 
\mbox{$\Omega_{\mathrm{K}}=0.000\pm0.005$} is the curvature density parameter and
\mbox{$\Omega_{m}=0.308\pm0.012$} is the density parameter of matter. This results in \mbox{$\Omega_\Lambda=0.692\pm0.017$}. Planck also demands that
the effective barotropic parameter of dark energy is 
\mbox{$w_{\mathrm{DE}}=-1.023^{+0.091}_{-0.096}$} (\textit{Planck} TT+lowP+ext)%
\footnote{``ext" includes the \textit{Planck} lensing, BAO, JLA and $H_0$ data 
sets.} at 2-$\sigma$. We investigate this in the appendix and we find that 
demanding that our model satisfies these observational requirements results in 
the bound \mbox{$\alpha\gtrsim 1.5$} (i.e. \mbox{$\sqrt{6\alpha}\gtrsim 3$}).   
In all cases, the scalar field has unfrozen but is yet to settle on the 
attractor solution. This results in $\dot{w}_{\mathrm{DE}} \neq 0$, which lies 
within current Planck bounds (see appendix) but can be potentially observable 
in the near future, where the dot denotes time derivative.

We can obtain an upper bound on $\alpha$ by avoiding super-Planckian values for
the non-canonical field $\phi$. The motivation for this is to suppress radiative
corrections and the 5th-force problem, which plague quintessence models
\cite{Dimopoulos:2017zvq}. However, the bound is soft, as both 
loop corrections and interactions are suppressed near the poles \cite{Linde:2017pwt,Kallosh:2016gqp} as 
we discuss in the penultimate section of this paper. 
Still, being conservative, we choose to avoid 
a super-Planckian non-canonical inflation field. Therefore, the relevant range 
for $\alpha$ is the following:
\begin{equation}\label{eq:alpha_range}
3 \lesssim \sqrt{6\alpha} \lesssim 5 \quad\Leftrightarrow\quad
1.5 \leq \alpha \leq 4.2
\,.
\end{equation}
For the above range, there is a theoretical prejudice in view of maximal supergravity, string theory, and M-theory, for particular values of $\alpha$ satisfying $3\alpha = 5,6,7 $ \cite{Kallosh2017,Ferrara:2016fwe,Kallosh2017a}.

\subsection{Inflationary Observables}

The inflationary observables predicted by this model are 
\cite{Dimopoulos:2017zvq}:
\begin{equation}\label{eq:r}
r = 16\epsilon = 12\alpha \left(  N_* + \frac{\sqrt{3\alpha}}{2} \right)  ^{-2}	\,,
\end{equation}
\begin{gather}\label{eq:n_s_final}
n_s  
= 1-\frac{2}{\Big(N_*+\frac{\sqrt{3\alpha}}{2}\Big)} - 
\frac{3\alpha}{2\Big(N_* + \frac{\sqrt{3\alpha}}{2}\Big)^2} 
\;\simeq\; 
1 - \frac{2}{N_*}	\,,
\end{gather}
and
\begin{multline}
n_s'\equiv\frac{{\rm d}\ln n_s}{{\rm d}\ln k} \\
=
-\frac{1}{\Big(N_*+\frac{\sqrt{3\alpha}}{2}\Big)}
\frac{2\Big(N_*+\frac{\sqrt{3\alpha}}{2}\Big)+3\alpha}
{\Big(N_*+\frac{\sqrt{3\alpha}}{2}\Big)^2-
2\Big(N_*+\frac{\sqrt{3\alpha}}{2}\Big)-\frac32\alpha} \\
\simeq
-\frac{2}{N_*^2-2N_*}
\label{nsrunning}
\end{multline}
where $r$ is the tensor to scalar ratio, $n_s$ is the spectral index of the 
scalar perturbations and $n_s'$ its running. In the above, 
$N_*$ is the number of remaining inflationary e-folds when the cosmological scales left the horizon 
during inflation and the last equations in Eqs.~\eqref{eq:n_s_final} and
\eqref{nsrunning} correspond 
to \mbox{$\alpha\ll N_*^2$} and are the standard $\alpha$-attractors results.

The value of $N_*$ is dependent on $T_{\mathrm{reh}}$ via the equation\footnote{%
In Ref.~\cite{Dimopoulos:2017zvq}, $N_*$ was calculated exactly to be 63.49, 
due to a cancellation of all dependence on $T_{\mathrm{reh}}$ in the case of 
gravitational reheating.} 
\begin{equation}\label{N*}
N_*\simeq 61.93 + \mathrm{ln}\Big(\frac{V_{\mathrm{end}}^{1/4}}{m_P}\Big) + 
\frac{1}{3}\mathrm{ln}\Big(\frac{V_{\mathrm{end}}^{1/4}}{T_{\mathrm{reh}}}\Big) \,.
\end{equation}
As will be discussed in the following sections, $T_{\mathrm{reh}}$ depends on $n$ 
and $\alpha$ as well as the efficiency of the reheating mechanism. As such, the
value of $N_*$ is determined iteratively. However, 
using $N_* = 62$ (in view of Eq.~\eqref{N*}) and \mbox{$\sqrt{6\alpha}=4$} (the 
middle point of the range in Eq.~\eqref{eq:alpha_range}) in 
Eqs.~\eqref{eq:n_s_final} and \eqref{nsrunning}
gives approximate results:
\begin{equation}\label{eq:ns_results}
n_s = 0.968\quad{\rm and}\quad n_s'=-5.46\times 10^{-4}\,.
\end{equation}
Assuming $N_* = 62$ and considering the range in Eq.~\eqref{eq:alpha_range}, 
Eq.~\eqref{eq:r} gives
\begin{equation}\label{eq:r_results}
0.005 \leq r \leq 0.012  \,.
\end{equation}
These inflationary observables are in 
excellent agreement with the latest Planck observations \cite{Ade:2015lrj}.
As shown later, the resulting values of $n_s$, $n_s'$ and $r$ using the actual 
values of $N_*$, obtained from considering the reheating mechanism and the 
quintessence requirements, remain in excellent agreement with the observations
and are very close to the above.

Finally, for the energy scale of inflation we find \cite{Dimopoulos:2017zvq},
\begin{equation}\label{eq:InflationaryScale}
\left(\frac{M}{m_P}\right)^2= \frac{3\pi\sqrt{2\alpha\mathscr{P}_{\zeta}}}{
\Big(N_*+\frac{\sqrt{3\alpha}}{2}\Big)} \, \mathrm{exp}
\left[\frac{3\alpha}{4}\Big(N_*+\frac{\sqrt{3\alpha}}{2}\Big)^{-1}\right]
\,,
\end{equation}
where \mbox{$\mathscr{P}_{\zeta} = (2.199 \pm 0.066) \times 10^{-9}$}, is the 
spectrum of the scalar curvature perturbation \cite{Ade:2015lrj}. With
\mbox{$N_*=62$} and $\alpha$ in the range in Eq.~\eqref{eq:alpha_range},
we find \mbox{$M\simeq 10^{16}\,$GeV}, which is at the scale of grand 
unification. The actual values of $M$ can be seen in Fig.~\ref{fig:M}.

\section{Reheating and Quintessence}

\subsection{Inflaton Freezing}

As mentioned earlier, $n$ affects the freezing value of the field, $\varphi_F$. During kination the field is oblivious of the potential and the Klein-Gordon equation reduces to $\ddot{\varphi} + 3H\dot{\varphi} \simeq 0$. Consequently, following the treatment in Ref.~\cite{Dimopoulos:2017zvq}, it is easy to show
that during kination, the scalar field grows as
\begin{equation}\label{phikin}
\varphi=\varphi_{\rm IP}+\sqrt{\frac23}\,\mpl\ln\left(\frac{t}{t_{\rm IP}}\right)
\,,
\end{equation}
where the subscript `IP' denotes the moment of instant preheating, when 
radiation is generated  (discussed in the following subsection), and we 
consider that kination continues for a while after instant preheating occurs. 
Because radiation density scales as \mbox{$\rho_r\propto a^{-4}$}, once created, 
radiation eventually takes over, since for a kinetically dominated scalar field 
we have \mbox{$\rho_{\rm kin} \equiv \frac{\dot{\varphi}^2}{2} \propto a^{-6}$}. Thus, for the density parameter of 
radiation during kination we have 
\mbox{$\Omega_r=\rho_r/\rho_{\phi}\propto a^{-2}$}.
Denoting as `reh' the moment of reheating, i.e. the moment when the radiation 
bath comes to dominate the Universe, we have \mbox{$\Omega_r^{\mathrm{reh}}=1$} by 
definition. Therefore, the radiation density parameter at instant preheating is
\begin{equation}\label{eq:Omega_in_t}
\Omega_r^{\mathrm{IP}}=
\Omega_r^{\mathrm{reh}}\Big(\frac{a_{\mathrm{IP}}}{a_{\mathrm{reh}}}\Big)^2= 
\Big(\frac{t_{\mathrm{IP}}}{t_{\mathrm{reh}}}\Big)^{\frac{2}{3}}\,,
\end{equation}
where $\Omega^{\mathrm{IP}}_r \equiv (\rho_r/\rho)_{\mathrm{IP}}$ is the radiation 
density parameter at instant preheating 
and we considered that during kination $a \propto t^{1/3}$. Inserting the 
above into Eq.~\eqref{phikin} we find
\begin{equation}
\varphi_{\rm reh}=\varphi_{\rm IP}-\sqrt{\frac32}\,\mpl\ln(\Omega_r^{\rm IP})\,.
\label{phireh}
\end{equation}
Now, as shown in Ref.~\cite{Dimopoulos:2017zvq}, during radiation domination,
the field continues to roll for a while as
\begin{equation}
\varphi=\varphi_{\rm reh}+
\sqrt{\frac23}\,\mpl\left(1-\sqrt{\frac{t_{\rm reh}}{t}}\right)\,.
\label{phiRD}
\end{equation}
The above suggests that the field freezes at a value $\varphi_F$, given by
\begin{equation}\label{eq:phi_F_from_Omega}
\varphi_F=\varphi_{\rm IP}+\sqrt{\frac23}
\Big(1-\frac32\,\ln\Omega^{\rm IP}_r\Big)\mpl \,,
\end{equation}
where we used Eq.~\eqref{phireh}. Here, we assume that the generation of
radiation is almost instantaneous, as is discussed later. 
A relationship between $n$ and 
$\varphi_F$ can be obtained from the final energy density requirements for dark 
energy. Starting from the requirement that the density of quintessence 
must be comparable to the density of the 
Universe today:
\begin{equation}\label{eq:phi_F_from_n_and_alpha}
\frac{\rho_{\mathrm{inf}}}{\rho_0} \simeq 
\frac{M^4}{V(\varphi_F)}
\simeq \frac{e^{2\varphi_F/\sqrt{6\alpha}\mpl}}{2ne^{-2n}} \simeq 10^{108} \,,
\end{equation}
where $\rho_{\mathrm{inf}}$ is the energy density during inflation, we find
\begin{equation}\label{eq:n_phi_F}
2n-\mathrm{ln}(2n)=
108\,\mathrm{ln}10-\frac{2}{\sqrt{6\alpha}}\frac{\varphi_{F}}{\mpl} \,,
\end{equation}
and combining Eqs.~\eqref{eq:phi_F_from_Omega} and \eqref{eq:n_phi_F} gives
\begin{equation}\label{}
2n-\ln(2n)=108\,\ln 10 - \frac{2}{\sqrt{6\alpha}}\sqrt{\frac23}
\Big(1 - \frac{3}{2}\,\mathrm{ln}\Omega_r^{\mathrm{IP}}\Big) \,,
\end{equation}
where we assumed that the right-hand-side of Eq.~\eqref{eq:phi_F_from_Omega}
is dominated by the last term. This is so when \mbox{$\Omega_r^{\rm IP}\ll 1$}, 
which can be challenged only for 
very high reheating efficiency. However, as we show later, such efficiency is
excluded because of backreaction constraints. 
Also, 
high reheating efficiency would mean that radiation domination begins almost
right after instant preheating. This would result in a high reheating 
temperature, incompatible with gravitino over-production considerations.

The parameter space for $n$ is related to the density of produced radiation at 
the end of inflation. Hence, changing the reheating efficiency affects the 
parameter space for~$n$. $\Omega_r^{\mathrm{IP}}$~is larger the more 
efficient instant preheating is, meaning the scalar field rolls less far in 
field space before it freezes. So, maintaining the same final energy density 
(comparable to the density at present), requires a higher $n$ value. 
%
%
To find bounds on $n$, we derive bounds on $\Omega_r^{\mathrm{IP}}$ and 
evolve the equations of motion 
numerically, in order to determine exactly when instant preheating occurs and 
how this affects the variables we need to constrain. 

The equations of motion used are: 
\begin{eqnarray}
3\mpl^2H^2 & = & \frac{1}{2}\dot{\varphi}^2 + V(\varphi)  \,, \\
-2\dot{H}\mpl^2 & = & \dot{\varphi}^2  \,, \\
\ddot{\varphi} & = &-3H\dot{\varphi} -V'(\varphi)  \,,
\end{eqnarray}
where the prime denotes differentiation with respect to $\varphi$ and dots denote differentiation with respect to time.
\pagebreak
\subsection{Instant Preheating}

For instant preheating we presume the inflaton $\phi$ is coupled to some other
scalar field $\chi$. In particular, we consider an interaction at an enhanced 
symmetry point (ESP) at $\phi=\phi_0$. The Lagrangian density near the ESP is
\begin{equation}\label{Lphi0}
\mathcal{L} = \mathcal{L}(\phi_0) + \mathcal{L}_{\mathrm{int}} \,,
\end{equation}
where ${\cal L}(\phi_0)$ is determined by Eq.~\eqref{L0} evaluated at 
$\phi_0$. The interaction Lagrangian density near the ESP is
\begin{equation}\label{IPL}
\mathcal{L}_{\mathrm{int}}=-\frac12 g^2(\phi-\phi_0)^2\chi^2-h\chi\psi\bar{\psi} \,,
\end{equation}
where $g$ and $h$ are perturbative coupling constants, and 
$\psi$ denotes a fermion field, coupled to $\chi$. The fermion is taken to be 
light, such that the $\chi$-particles decay into a radiation bath. We consider
\mbox{$h\sim 1$}, which means that the decay of $\chi$ is immediate.

The scalar field, $\chi$, can be expressed in terms of the creation and 
annihilation operators, and the Fourier modes of this expansion obey a wave 
equation with a frequency dependent on the effective mass of $\chi$. Certain 
solutions to this wave equation are growing solutions and this translates into 
an exponential increase of the occupation number $n_k$ for a particular mode, 
when particle production occurs \cite{Felder:1998vq,Campos:2002yk}. The adiabaticity condition 
\begin{equation}
\frac{\dot{\omega}_k}{\omega^2_k} < 1 \,,
\label{adiabaticityparameter}
\end{equation}
where $\omega_k$ is the frequency of the Fourier expanded wave equation, must 
be violated for particle production to occur. For the interaction terms used 
here, this leads to:
\begin{equation}\label{eq:flat_adiabaticity}
|\dot{m}_{\chi}|\ll m_{\chi}^2 \,,
\end{equation}
with \mbox{$m_{\chi}^2 = g^2(\phi-\phi_0)^2$}. Thus, particle production takes 
place when:
\begin{equation}\label{adi_cond_shift}
|\dot{\phi}| > g(\phi-\phi_0)^2 \,,
\end{equation}
which gives the following range for $\phi$
\begin{equation}\label{eq:PParea}
\phi_0 - \sqrt{\frac{|\dot{\phi}|}{g}} \leq \phi \leq \phi_0 + \sqrt{\frac{|\dot{\phi}|}{g}} \,.
\end{equation}
The above is the window of $\phi$ in which particle production occurs.

\begin{figure}
	\centering
	\includegraphics[width=1\linewidth]{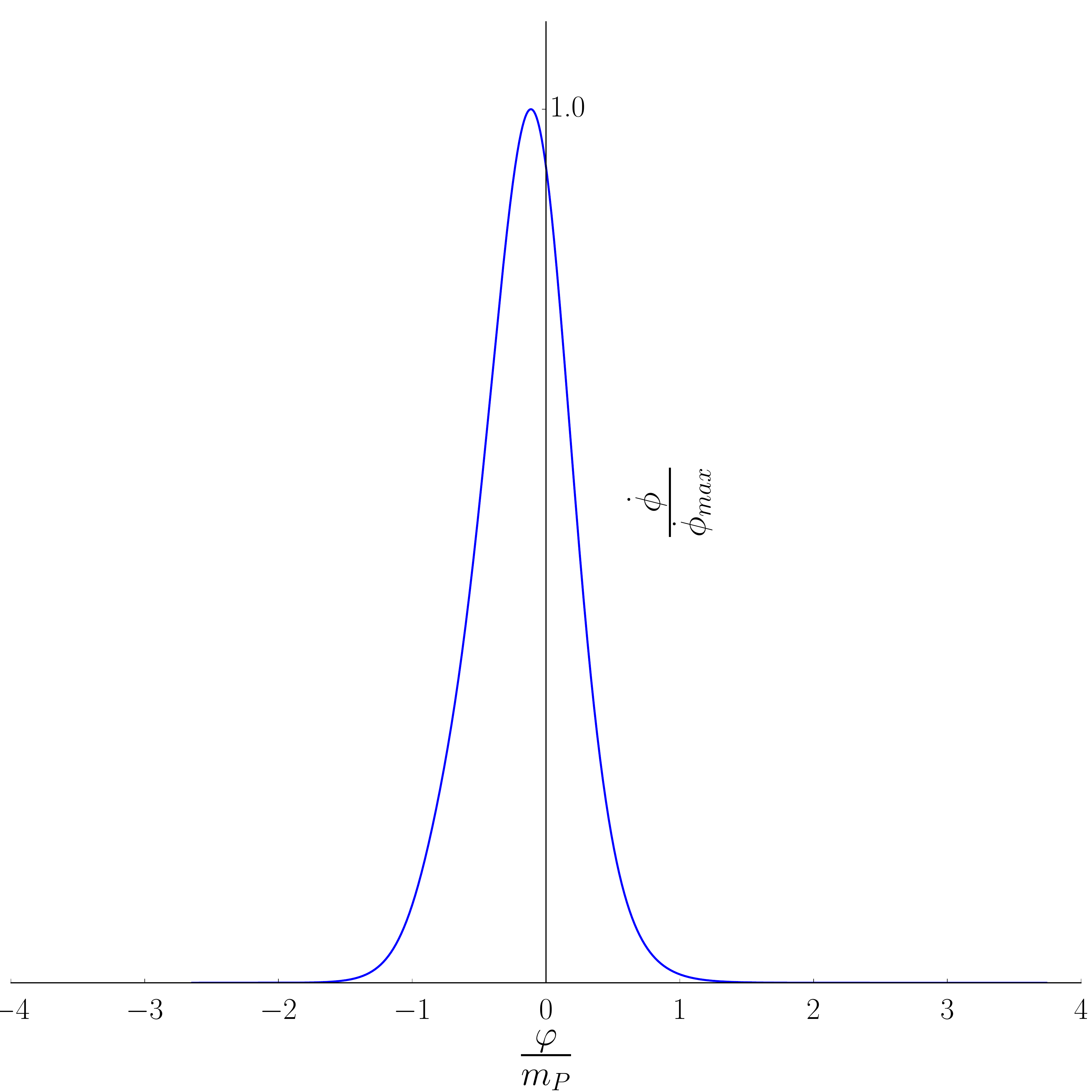}
\caption{This plot depicts where $\dot{\phi}$ is maximised in the $\varphi$ 
direction.}
	\label{fig:pert_validation}
\end{figure}

The careful reader may have noticed that the interaction considered regards 
$\phi$, the original non-canonically normalised field.
Hence, we need to find ${\phi}$ and $\dot{{\phi}}$ to check the adiabaticity 
constraint. 
We can find $\phi$ using Eq. \eqref{eq:canonicalVariables} from which we 
readily obtain
\begin{equation}\label{eq:phidot_in_tilde}
\dot{\phi} = 
\mathrm{sech}^2\Big(\frac{\varphi}{\sqrt{6\alpha}\mpl}\Big)\dot{\varphi} \,,
\end{equation}
where we obtain $\varphi$ and $\dot{\varphi}$ from the computation, but for 
completeness:
\begin{align}
\label{eq:tildes_in_phi}
\varphi = \sqrt{6\alpha}\mpl\,\mathrm{tanh}^{-1}
\Big(\frac{\phi}{\sqrt{6\alpha}\mpl}\Big) \\
\mathrm{and} \qquad \dot{\varphi} = \frac{\dot{\phi}}{1 - \frac{\phi^2}{6\alpha\mpl^2}} \,,
\end{align}
which are analytically cyclic. 
It is clear from this computation that the region where $\dot{\phi}$ is 
maximised and particle production occurs is very close to $\phi=0$, meaning 
$\phi \simeq \varphi$ (c.f. Eq.~\eqref{eq:canonicalVariables}) and $\phi$ is 
almost canonical.
This can be seen clearly in Fig.~\ref{fig:pert_validation}.
This is because, when the non-canonical $\phi$ is near the poles it hardly 
varies, even when the canonical $\varphi$ changes substantially. Thus, it is 
not possible to violate the adiabaticity condition in 
Eq.~\eqref{adiabaticityparameter} in this region. Therefore, there may be many 
ESPs along the $\varphi$ direction, but only near 
\mbox{$\phi\simeq\varphi\simeq 0$} can we have particle production. 

The number density of produced $\chi$ particles \cite{Felder:1998vq,Campos:2002yk} is
\begin{equation}\label{eq:number_density_chis}
n_{\chi}=\int\frac{\mathrm{d}^3k}{(2\pi)^3}\,n_k= 
\frac{1}{2\pi^2}\int_{0}^{\infty}k^2n_k\mathrm{d}k\,,
\end{equation}
where the occupation number 
\begin{equation}\label{key}
n_k = \mathrm{exp}\Big(-\frac{\pi m_{\chi}^2}{\dot m_{\chi}}\Big) \,,
\end{equation}
is suppressed when $\dot m_{\chi} < m_{\chi}$, evidencing why the adiabaticity 
condition in Eq.~\eqref{eq:flat_adiabaticity}, controls particle production. 
Combining Eq.~\eqref{eq:number_density_chis} with the $\chi$ particle effective
mass provides the density of the produced $\chi$ particles \cite{Felder:1998vq,Campos:2002yk} 
\begin{equation}\label{eq:rho_density_1}
\rho_{\chi}^{\rm IP}=
\frac{g^{5/2}|\dot{\phi}_{\mathrm{IP}}|^{3/2}\phi_{\mathrm{IP}}}{8\pi^3}\,.
\end{equation}

The instant preheating efficiency is maximised when $\phi$ is near the final
edge of the production window in Eq.~\eqref{eq:PParea} because, even though we 
expect a continuous contribution to $n_{\chi}$ whilst $\phi$ is in this region, 
the produced $\chi$-particles are diluted by the expansion of the Universe. 
Therefore, we expect only the ones produced near the end of the particle production regime to contribute significantly to $\rho_{\chi}^{\rm IP}$.
As such, from Eq.~\eqref{eq:PParea}, taking \mbox{$\phi_0\simeq 0$}, we set
\begin{equation}\label{eq:pip_extremum_value}
\pip = \sqrt{\frac{\dot{\phi}_{\mathrm{IP}}}{g}} \,,
\end{equation} 
which simplifies Eq.~\eqref{eq:rho_density_1} to
\begin{equation}\label{eq:rho_density_2}
\rho_r^{\rm IP}=\rho_{\chi}^{\rm IP}=\frac{g^{2}\dot{\phi}_{\mathrm{IP}}^{2}}{8\pi^3}\,,
\end{equation}
where we have considered that $\dot{\phi}>0$ because the field is rolling 
towards larger values and we have assumed that the decay of the $\chi$-particles to 
radiation is instantaneous.

For each choice of $n$, the quintessence requirements stipulate the required 
value of $\varphi_F$ and hence $\Omega_r^{\mathrm{IP}}$. 
The value of $\Omega_r^{\mathrm{IP}}$ is 
\begin{equation}\label{key}
\Omega_r^{\mathrm{IP}} = 
\frac{\rho_{\chi}^{\rm IP}}
{\rho_{\chi}^{\rm IP} + \rho_{\phi,a}^{\rm IP}}=
\frac{\rho_r^{\rm IP}}{\rho_{\phi,b}^{\rm IP}}
\,.
\end{equation}
where \mbox{$\rho_{\chi}^{\rm IP}=\rho_r^{\rm IP}$} is defined in 
Eq.~\eqref{eq:rho_density_2} and 
the subscript~`$a/b$' refers to after/before instant preheating.
%
%
Inserting the above in
Eq.~\eqref{eq:rho_density_2}, a rearrangement quickly yields:
\begin{equation}\label{eq:g_from_Omega}
g = \sqrt{\frac{8\pi^3}{\dot\phi_{\rm IP}^2}\,\Omega_r^{\mathrm{IP}}\rho_\phi^{\rm IP}
} \,,
\end{equation}
where we have omitted subscript `$b$' for simplicity. Note that 
\mbox{$\rho_{\phi,a}\simeq\rho_{\phi,b}$} when \mbox{$\Omega_r^{\rm IP}\ll 1$}.

For each choice of $n$, we calculate $\varphi_F$ from 
Eq.~\eqref{eq:phi_F_from_n_and_alpha} and insert this into 
Eq.~\eqref{eq:phi_F_from_Omega} to obtain $\Omega_r^{\mathrm{IP}}$ as a 
function of $n$. As the reheating variables are also functions of $n$, we now 
have $g$ in terms of only $n$. However, as noted previously, $\pip$ and 
$\dot{\phi}_{\mathrm{IP}}$ are themselves dependent on $g$ and so this
requires iteration. This is the procedure to obtain a value of $g$ for a given 
value of $n$.
\pagebreak
\section{Constraints from Reheating and Quintessence}

\subsection{\boldmath Immediate Constraints on $n$}

An immediate sanity check arises: if 
\mbox{$\varphi_{F} < \varphi_{{\mathrm{IP}}}$} then the combination of $n$ and 
$\alpha$ is disallowed. This allows us at first glance to constrain $n$. For 
the complete range of allowed $\alpha$ values, \mbox{$1.5\leq\alpha\leq 4.2$},
we find
\begin{equation}\label{key}
n \leq 130 \,.
\end{equation}
The fact that this approach produces an upper limit on $n$ makes sense because 
a larger $n$ value makes the potential steeper and means lower $V$ values will 
be reached earlier in field space. Hence, to equate $V(\varphi_F)$ with dark 
energy today will require a lower value for $\varphi_F$. As such, ensuring 
\mbox{$\varphi_F > \varphi_{{\mathrm{IP}}}$} results in an upper bound on $n$. 

\subsection{\boldmath Keeping $g$ Perturbative and Ensuring Radiation Domination}

The first constraint on $g$ is found by requiring $g<1$, for a perturbative 
coupling constant, which provides a tight upper bound on $n$:
\begin{align}\label{eq:n_g_upper_bound_1}
\alpha = 1.5:  \qquad n \leq 124 \,, \nn \\
\alpha = 4.2:  \qquad n \leq 125 \,.
\end{align}

However, to obtain the correct Universe history, we also need to ensure we have 
a period of radiation domination after instant preheating, which might provide 
a tighter bound. In a quintessential inflation model with a period of kination 
after radiation generation, this is never a problem because the density of the 
produced radiation scales as \mbox{$\rho_r\propto a^{-4}$} whilst the density of 
the kinetically dominated field scales as \mbox{$\rho_\phi\propto a^{-6}$}. 
Hence, to ensure radiation domination we need to ensure that the scalar field 
remains kinetically dominated after instant preheating. Note, that the transfer 
of energy to $\chi$-particles during instant preheating comes from the kinetic 
energy density of the inflaton only, 
therefore \mbox{$V(\phi_a)=V(\phi_b)\equiv V(\pip)$}. Were there not 
enough kinetic density left, the inflaton would become potentially dominated 
and would embark to a new bout of inflation. Thus, we need to ensure that the kinetic 
energy of the inflaton is greater than the potential energy after instant 
preheating. This leads to
\begin{equation}\label{eq:IPlimit}
\rho_{\phi,a} - V(\phi_{\mathrm{IP}}) > V(\phi_{\mathrm{IP}}) \quad 
\Rightarrow
\quad \rho_{\chi} < \rho_{\phi,b} - 2 V(\phi_{\mathrm{IP}}) \, ,
\end{equation} 
where the subscripts `$a$' and `$b$' refer to after and before instant 
preheating respectively and we have used that 
\mbox{$\rho_{\phi,b}=\rho_{\phi,a}+\rho_\chi$}. Eq.~\eqref{eq:IPlimit} gives us an 
upper limit on the allowed energy density of produced $\chi$ particles, which 
translates to an upper limit on the perturbative coupling $g$, from the equation
for the energy density, Eq.~\eqref{eq:rho_density_2}. However, it turns out that
this constraint is automatically satisfied for a perturbative coupling with 
\mbox{$g<1$}.

\begin{table}[hb]
	\begin{center}
		\begin{tabular}{|c|c|c|}
			\hline
			$g$ & Allowed $n$ values & Allowed $\kappa$ values \\
			\hline
			\hline
			0.001 & $119 \leq n \leq 122$ & $24.3 \leq \kappa \leq 39.6$ \\		
			\hline
			0.01 & $121 \leq n \leq 123$  & $24.5 \leq \kappa \leq  40.3$ \\
			\hline
			0.1 & $123 \leq n \leq 124$ & $24.7 \leq \kappa \leq  41.0$	\\
			\hline
			1.0 & $125 \leq n \leq 126$ & $25.1 \leq \kappa \leq  41.7$	\\
			\hline
		\end{tabular}
		\caption{Allowed $n$ and $\kappa$ values for specific choices of $g$, within the allowed $\alpha$ range, before consideration of backreaction and gravitino constraints}
		\label{tab:results_for_g}	
	\end{center}
\end{table}

\begin{table}[h]
	\begin{center}
		\def\arraystretch{1.3}
		\begin{tabular}{|c|c|c|c|c|c|c|}
			\hline
			$\alpha$ & $n$ & $N_*$ & $n_s$ & $r / 10^{-3}$         & $n_s' / 10^{-4}$ \\
			\hline
			\hline
			1.5 	 & 118 & 62.7  & 0.968	& $4.42$ & $-5.25$ \\		
			\hline
			1.5 	 & 124 & 59.1  & 0.966 & $4.97$ & $-5.92$ \\
			\hline
			\hline
			4.2      & 121 & 63.5  & 0.968	& $11.8$ & $-5.11$ \\
			\hline
			4.2      & 125 & 59.4  & 0.966	& $13.9$ & $-5.86$ \\
			\hline
		\end{tabular}
		\caption{For the allowed range of $n$, prior to consideration of backreaction and gravitino constraints, the corresponding values of $N_*$ and the inflationary observables are shown.}
		\label{tab:ns_r_results}	
	\end{center}
\end{table}

\begin{table}[h]
	\begin{center}
		\def\arraystretch{1.5}
		\begin{tabular}{|c|c|c|c|c|c|c|}
			\hline
			$\alpha$ & $n$ & $\kappa$ & $T_{\mathrm{reh}}$ (GeV)      & $M$ (GeV)& $V_0^{1/4}$ (GeV)    \\
			\hline
			\hline
			1.5 	 & 118 & 39.3	  & $3.84\times10^{6}$ & $8.50\times 10^{15}$	& $1.31\times 10^{3}$ \\		
			\hline
			1.5 	 & 124 & 41.3	  & $2.22\times10^{11}$ & $8.76\times 10^{15}$ 	& $3.01\times 10^{2}$ \\
			\hline
			\hline
			4.2      & 121 & 24.1	  & $2.35\times10^{5}$ & $1.10\times 10^{16}$	& $8.04\times 10^{2}$ \\
			\hline
			4.2      & 125 & 24.9	  & $6.07\times10^{10}$ & $1.15\times 10^{16}$  & $3.06\times 10^{2}$ \\
			\hline
		\end{tabular}
		\caption{For the allowed range of $n$, prior to consideration of backreaction and gravitino constraints, the corresponding values of $T_{\mathrm{reh}}$, $M$ and $V_0^{1/4}$ are shown.}
		\label{tab:Energy_scale_results}	
	\end{center}
\end{table}

\subsection{Backreaction Constraint}

We must also consider the back reaction of produced $\chi$-particles on $\phi$, 
which may further constrain the allowed value of~$g$. The equation of motion for
the scalar field, including back reaction, is given by \cite{Felder:1998vq,Campos:2002yk}
\begin{equation}\label{eq:EoM_backreaction}
\ddot{\phi} + 3H\dot{\phi} + V'(\phi) = -gn_{\chi}\frac{\phi}{|\phi|} \,,
\end{equation}
where 
\begin{equation}\label{key}
n_{\chi} = \frac{(g|\dot{\phi}|)^{3/2}}{8\pi^3}\,
\exp\Big(-\frac{\pi m_{\chi}^2}{\dot{m}_{\chi}}\Big) 
\end{equation}
and we consider that, near the ESP, $\phi$ is canonically normalised 
(\mbox{$\phi\simeq\varphi$}), as discussed.

The exponential is suppressed during particle production and so the right hand 
side of Eq.~\eqref{eq:EoM_backreaction} becomes
\begin{equation}\label{key}
\ddot{\phi}+3H\dot{\phi} + V'(\phi) = -\frac{g^{5/2}\dot{\phi}^{3/2}}{8\pi^3} \,,
\end{equation}
where we have also considered $\dot\phi>0$. As back reaction increases, the 
magnitude of the right-hand side of this equation grows to have more and more 
of an effect on the dynamics \cite{Campos:2002yk}. This is maximised at
\mbox{$\phi=\phi_{\rm IP}$} (i.e. for maximum $n_\chi$). Computing this at that 
moment, we find that to avoid back-reaction effects requires roughly 
\begin{equation}
g \lesssim 10^{-3}\,.
\end{equation} 
In detail, the above upper bound on $g$ depends on the value of $\alpha$ as 
depicted in our results, see Figs.~\ref{fig:n_results_for_g}-\ref{fig:Lambda}.

\subsection{Gravitino Constraint}

Finally, because this is a model rooted in supergravity, constraints from 
over-production of gravitinos have to be taken into account. 
The over-production of gravitinos needs to be controlled because they can either
contribute to the mass of dark matter and overclose the Universe or they can 
decay and disrupt the production of nuclei during BBN. Gravitino production is 
strongly correlated with reheating temperature.
In general, the bound 
\mbox{$T_{\rm reh}<\mathcal{O}(10^9)\,$GeV}
\cite{Ellis:1982yb,Kawasaki:2006hm,Kawasaki:2006gs} is adequate.\footnote{%
However, in some cases, the bound can be much a tighter: 
\mbox{$T_{\mathrm{reh}}< \mathcal{O}(10^6)\,$GeV} \cite{Kohri:2005wn}.}

We can derive $T_{\mathrm{reh}}$ in this model from the relationship
\begin{equation}\label{eq:T_reh_1}
T_{\mathrm{reh}} = \Big(\frac{30}{\pi^2g_*}\rho_r^{\mathrm{reh}}\Big)^{1/4}\,,
\end{equation}
where `reh' denotes the moment of reheating, which is the onset of the 
radiation era. Employing Eq.~\eqref{eq:Omega_in_t} we readily find
\begin{equation}\label{key}
\rho_r^{\rm reh}=\rho_{\phi}^{\mathrm{reh}} 
=\rho_{\phi}^{\mathrm{IP}}(\Omega_r^{\mathrm{IP}})^3\,,
\end{equation}
where we used that \mbox{$\rho_{\phi} \propto a^{-6}$} during kination.
Inserting this into Eq.\eqref{eq:T_reh_1} we find
\begin{equation}\label{Treh}
T_{\mathrm{reh}} = 
\Big[\frac{30}{\pi^2g_*}\rho_\phi^{\rm IP}(\Omega_r^{\rm IP})^3\Big]^{1/4}
=\Big[\frac{30}{\pi^2g_*}\rho_r^{\rm IP} (\Omega_r^{\mathrm{IP}})^2\Big]^{1/4}
\,,
\end{equation}
where we also considered that \mbox{$\rho_r=\Omega_r\rho_\phi$}. We find 
$\Omega_r^{\rm IP}$ as follows
\begin{equation}
\Omega_r^{\rm IP}=\frac{\rho_r^{\rm IP}}{\rho_\phi^{\rm IP}}=
\frac{g^2\dot\phi_{\rm IP}^2}{8\pi^3}\frac{2}{\phi_{\rm IP}^2}=\frac{g^2}{4\pi^3}
\,,
\label{OIP}
\end{equation}
where we considered Eq.~\eqref{eq:rho_density_2} and that 
\mbox{$\rho_\phi^{\rm IP}=\frac12\dot\phi_{\rm IP}^2$} during kination.
Thus, \mbox{$\Omega_r^{\rm IP}\sim 10^{-2}g^2$}, which means that, since
\mbox{$g<1$}, $\Omega_r^{\rm IP}$ is very small. Given that the dependence of 
$(\rho_\phi^{\rm IP})^{1/4}$ on $g$ is weak, Eq.~\eqref{Treh} suggests 
\mbox{$T_{\mathrm{reh}}\propto g^{3/2}$}.
This is easy to understand by considering that a large value of $g$ means that
more radiation is generated at instant preheating. Consequently, reheating 
happens earlier and therefore $T_{\rm reh}$ is large. To limit $T_{\rm reh}$ 
to small enough values we need to avoid a large $g$.

In our model, the bound \mbox{$T_{\mathrm{reh}}<\mathcal{O}(10^9)\,$GeV} translates
to an upper bound on $g$ of roughly
\begin{equation}\label{eq:g_gravitino_bound}
g \lesssim 10^{-2} \,.
\end{equation}
As in the previous subsection, in detail, the above upper bound on $g$ depends 
on the value of $\alpha$ as depicted in our results, see Figs.~\ref{fig:n_results_for_g}-\ref{fig:Lambda}.

\subsection{\boldmath A Lower Bound on $g$}

The first constraint on a lower $g$ value is to ensure that radiation domination
occurs before BBN, but this constraint is not a worry for we find 
\mbox{$T_{\rm reh}\gg 1\,$MeV} in all cases.

We may obtain a lower bound on $\rho_{\chi}^{\rm IP}$, and hence $g$, from the 
nucleosynthesis constraint on the energy density of produced gravitational waves
during kination. We follow the treatment in Ref.~\cite{Agarwal:2017wxo} to find
the lower bound on $g$. The BBN constraint demands
\begin{equation}\label{key}
\Big(\frac{\rho_g}{\rho_r}\Big)_{\mathrm{reh}} \lesssim 10^{-2}\,,
\end{equation}
where
\begin{equation}\label{key}
\Big(\frac{\rho_g}{\rho_r}\Big)_{\mathrm{reh}} = 
\frac{64}{3\pi}h_{\rm{GW}}^2\Big(\frac{\rho_{\phi}}{\rho_r}\Big)_{\rm{IP}}\,.
\end{equation}
Using the relations
\begin{equation}\label{key}
h_{\mathrm{GW}}^2 = \frac{H_{\mathrm{end}}^2}{8\mpl^2}
\quad{\rm and}\quad 
H_{\mathrm{end}}^2 \simeq \frac{V_{\mathrm{end}}}{3\mpl^2}\,,
\end{equation}
where the subscript `end' signifies the end of inflation, we can re-express 
this as
\begin{equation}\label{key}
\Big(\frac{\rho_g}{\rho_{\gamma}}\Big)_{\mathrm{reh}} = 
\frac{8}{9\pi}\frac{V_{\mathrm{end}}}{\mpl^4}
\frac{1}{\Omega_r^{\rm IP}}
\lesssim 10^{-2} \,.
\end{equation}
Substituting the above in Eq.~\eqref{OIP} we get
\begin{equation}\label{eq:lowest_g_full}
g \geq 20\pi\sqrt{\frac89}\frac{V_{\rm end}^{1/2}}{\mpl^2}
\simeq 10\Big(\frac{M}{\mpl}\Big)^2\sim 10^{-4}
\end{equation}
where we considered that \mbox{$V_{\rm end}=M^4 e^{-\sqrt{3\alpha}}$} 
\cite{Dimopoulos:2017zvq}. For the last equation we considered
\mbox{$M\simeq 10^{16}\,$GeV} as suggested by Fig.~\ref{fig:M}.

%

\begin{table}[!h]
	\begin{center}
		\def\arraystretch{1.3}
		\begin{tabular}{|c|c|c|c|c|c|}
			\hline
			$\alpha$ & $n$ & $N_*$ & $n_s$ & $r/10^{-3}$ & $n_s' /10^{-4}$\\
			\hline
			\hline
			1.5 	 & 118 & 62.74 & 0.968 & 4.42 & -5.25 \\		
			\hline
			1.5 	 & 119 & 62.14 & 0.968 & 4.51 & -5.35\\
			\hline
			\hline
			4.2      & 121 & 63.54 & 0.968 & 11.8 & -5.11\\
			\hline
			4.2      & 122 & 62.53 & 0.967 & 12.2 & -5.28\\
			\hline
		\end{tabular}
		\caption{Final values for the parameters when considering the tightest constraints on $g$, from the backreaction bound.}
		\label{tab:ns_r_results_tight}	
	\end{center}
\end{table}

\begin{table}[!h]
	\begin{center}
		\def\arraystretch{1.5}
		\begin{tabular}{|c|c|c|c|c|c|}
			\hline
			$\alpha$ & $n$ & $\kappa$ & $T_{\mathrm{reh}}$ (GeV)  & $M$ (GeV)& $V_0^{1/4}$ (GeV)\\
			\hline
			\hline
			1.5 	 & 118 & 39.3  & $3.84\times 10^{6}$  & $8.5\times 10^{15}$ & $1.31\times 10^{3}$ \\		
			\hline
			1.5 	 & 119 & 39.7 & $2.39\times 10^{7}$  &$8.5\times 10^{15}$ & $1.03\times 10^{3}$ \\
			\hline
			\hline
			4.2      & 121 & 24.1 & $2.35\times 10^{5}$  &$1.1\times 10^{16}$ & $8.04\times 10^{2}$ \\
			\hline
			4.2      & 122 & 24.3 & $5.07\times 10^{6}$  &$1.1\times 10^{16}$ & $6.31\times 10^{2}$ \\
			\hline
		\end{tabular}
		\caption{Final values for the parameters when considering the tightest constraints on $g$, from the backreaction bound.}
		\label{tab:ns_r_results_tight_energy}	
	\end{center}
\end{table}
\pagebreak
\section{Results and Discussion}

The two unavoidable constraints are the upper bound on $n$, ensuring $g<1$ 
because of perturbativity, and the lower limit on $g$ ensuring a 
period of kination that does not disturb BBN through overproduction of 
gravitational waves. This bound is \mbox{$g\gtrsim 10^{-4}$}.
These bounds result in the parameter space
\begin{equation}\label{eq:n_large_param_space}
\alpha = 1.5:  \qquad 118 \leq n \leq 124 \,,
\end{equation}
\begin{equation}\label{key}
\alpha = 4.2: \qquad 121 \leq n \leq 125 \,.
\end{equation}

\begin{figure}[t]
	\centering
	\includegraphics[width=1\linewidth]{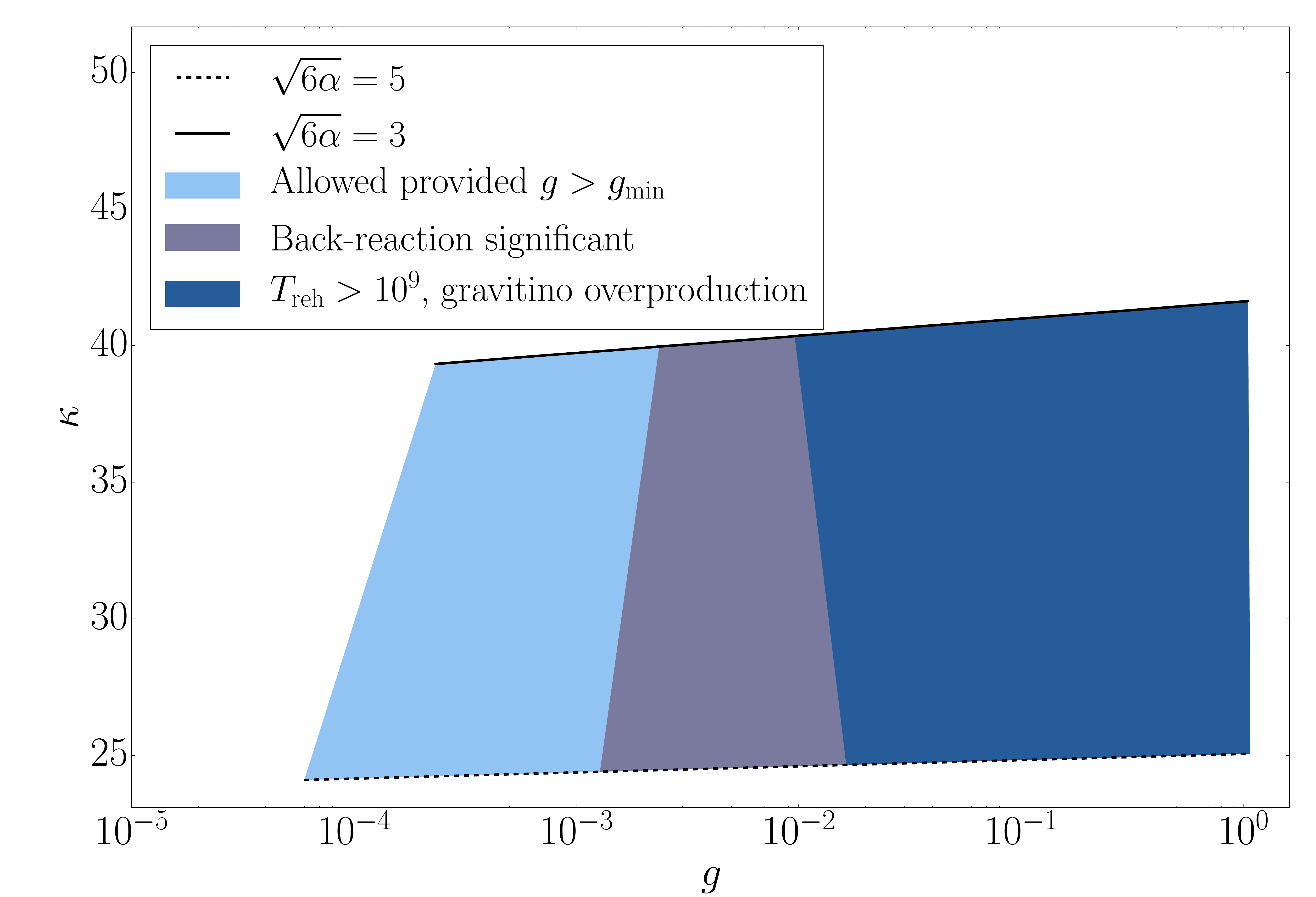}
	\caption{Allowed parameter space for $\kappa$, for the range of allowed $g$ values and allowed $\alpha$ values between 1.5 and 4.2. The bounds arising from backreaction and gravitino constraints are indicated.}
	\label{fig:kappa}
\end{figure}

\begin{figure}[t]
	\centering
	\includegraphics[width=1\linewidth]{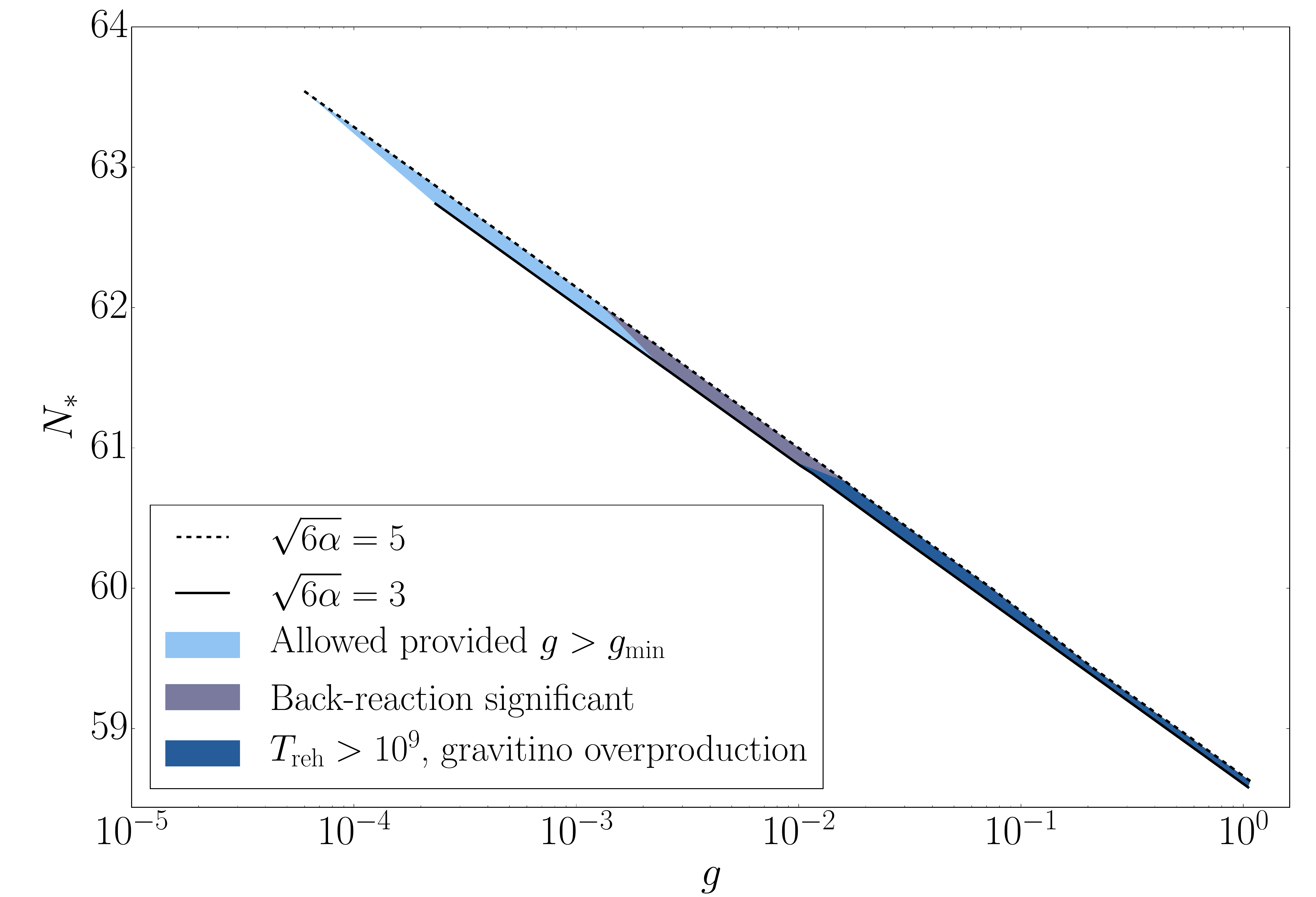}
	\caption{$N_*$ values for the range of $g$ values and allowed $\alpha$ values between 1.5 and 4.2. The bounds arising from backreaction and gravitino constraints are indicated.}
	\label{fig:Nstar}
\end{figure}

\begin{figure}[t]
	\centering
	\includegraphics[width=1\linewidth]{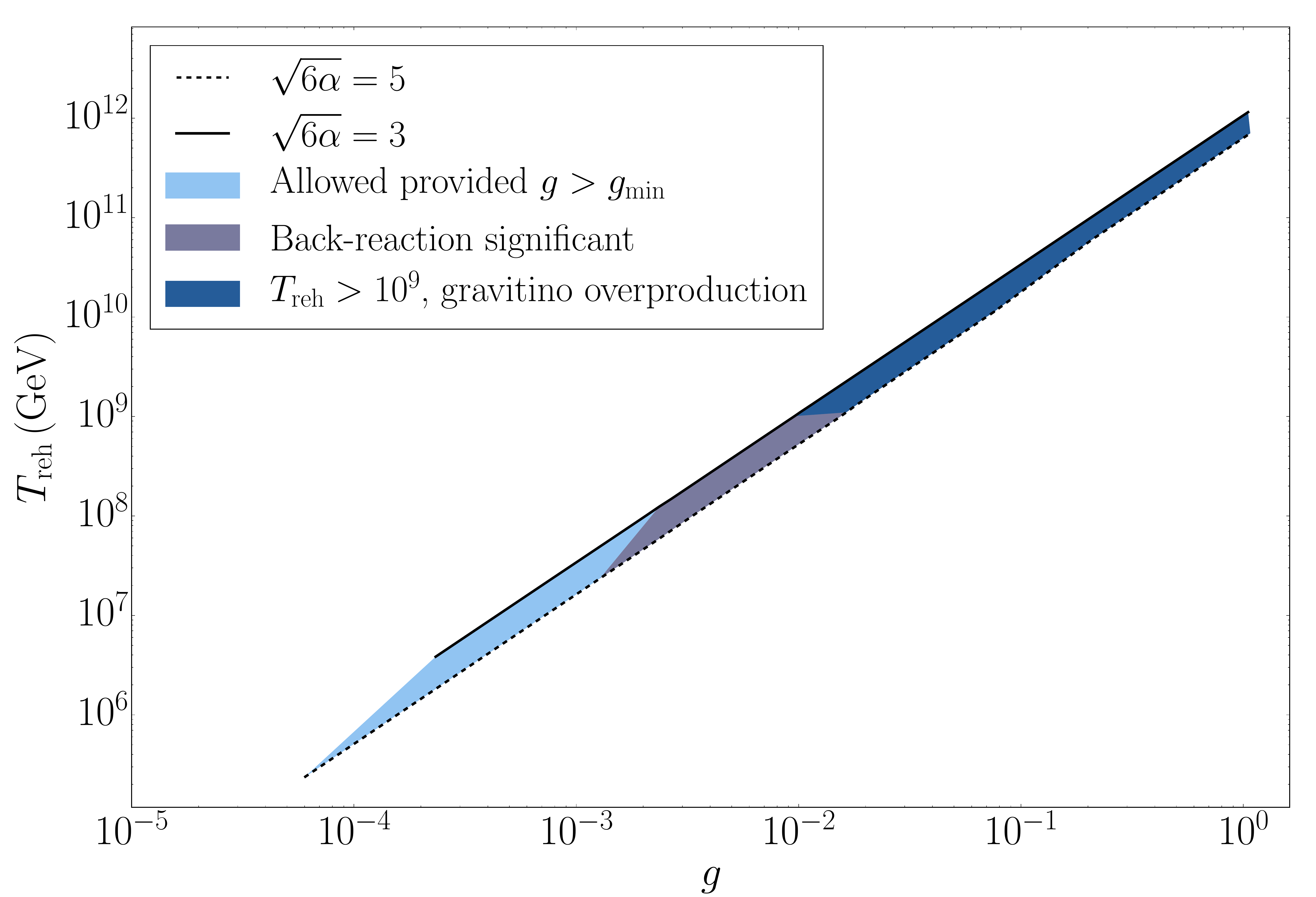}
	\caption{$T_{\mathrm{reh}}$ values for the range of $g$ values and allowed $\alpha$ values between 1.5 and 4.2. The bounds arising from backreaction and gravitino constraints are indicated.}
	\label{fig:Treh}
\end{figure}

The upper constraint on $g$ arising from the avoidance of backreaction in the 
instant preheating mechanism results in a bound on $g$ of approximately 
\mbox{$g \lesssim 10^{-3}$}. However, this bound can be sidestepped if the decay 
$\chi \rightarrow \psi\bar{\psi}$ is rapid, as is often assumed. All that 
is required is a large enough $h$ value for this coupling. 
\begin{figure}[h]
	\centering
	\includegraphics[width=1\linewidth]{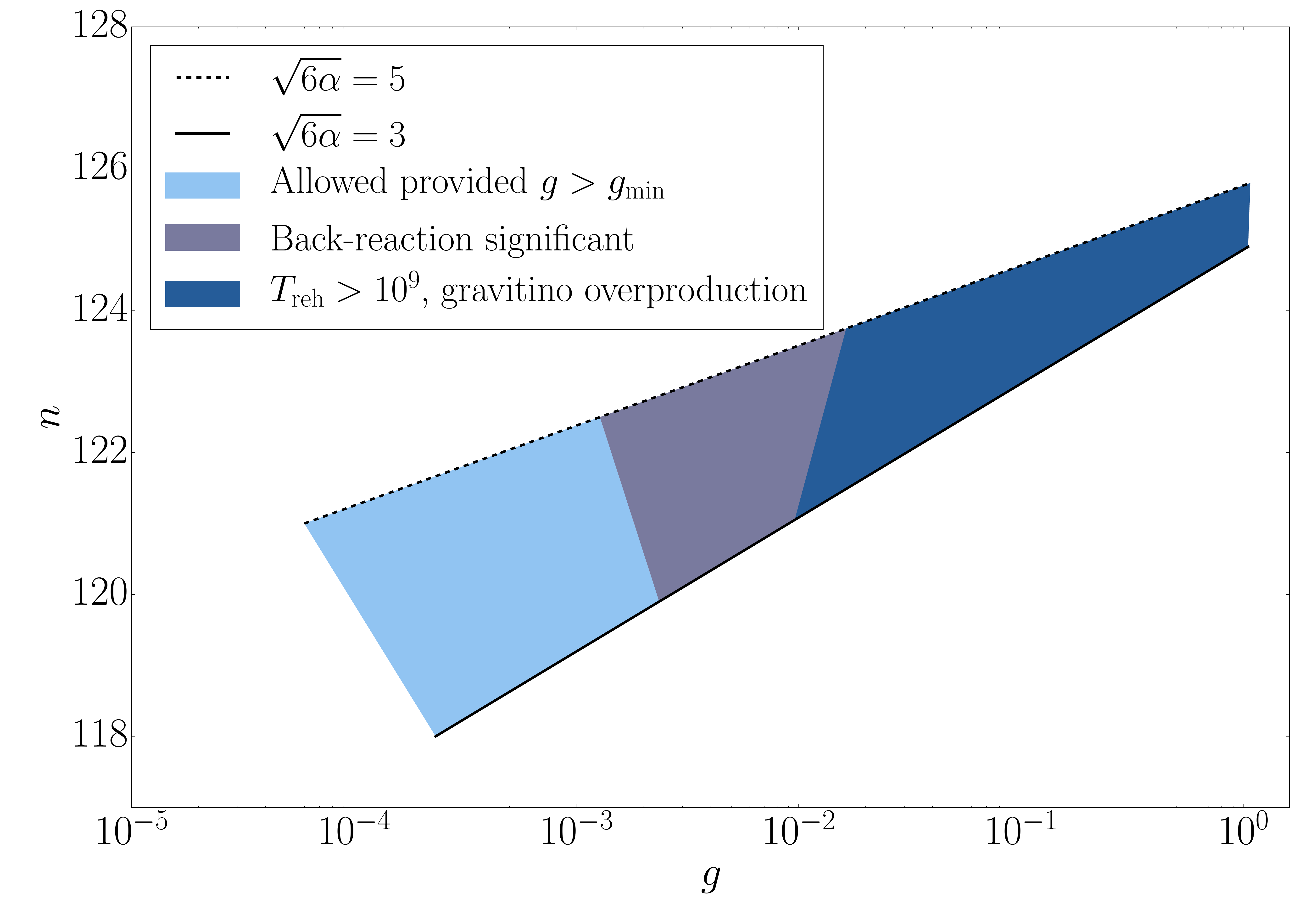}
	\caption{Allowed parameter space for $n$, for the range of allowed $g$ values and allowed $\alpha$ values between 1.5 and 4.2. The bounds arising from backreaction and gravitino constraints are indicated.}
	\label{fig:n_results_for_g}
\end{figure}
The upper bound on $g$ arising from gravitino over-production constraints
is important in a 
model rooted in supergravity. This bound is roughly \mbox{$g\lesssim 10^{-2}$}.
Because of this bound, the parameter space is reduced to
\begin{equation}\label{key}
\alpha = 1.5:  \qquad 118 \leq n \leq 122  \,,
\end{equation} 
\begin{equation}\label{key}
\alpha = 4.2: \qquad 121 \leq n \leq 124 \,.
\end{equation}
The allowed values of $n$ and $\kappa$ for a selection of $g$ values are shown 
in Table \ref{tab:results_for_g}, without consideration of the back-reaction and
gravitino bounds. For the extremal values of $n$ the corresponding values of 
$N_*$, $n_s$, $r$, $n_s'$, $T_{\mathrm{reh}}$, $M$ and $V_0^{1/4}$ are shown in Tables~\ref{tab:ns_r_results} and \ref{tab:Energy_scale_results}.
Figs. \ref{fig:n_results_for_g}, \ref{fig:kappa}, 
\ref{fig:Nstar}, \ref{fig:Treh}, \ref{fig:ns_r}, \ref{fig:M} \ref{fig:V0} and \ref{fig:Lambda} 
document how the parameter space is altered when the backreaction and gravitino 
bounds are included. Values of $N_*$, $n_s$ $r$, $n_s'$, $T_{\mathrm{reh}}$, $M$ and $V_0^{1/4}$ for the 
most constricted final parameter space are shown in 
Tables~\ref{tab:ns_r_results_tight} and \ref{tab:ns_r_results_tight_energy}. 

With a lower value of $\Omega_r^{\mathrm{IR}}$, the inflaton rolls to larger 
distances before it freezes. To fulfil dark energy requirements, this requires 
a lower $n$ value. The results found here for $n$ demonstrate this. 
The two different $\alpha$ values result in different $n$ requirements because 
$\alpha$ controls the slope of the quintessential tail 
(c.f. Eq.~\eqref{eq:V_quint}). A smaller/larger $\alpha$-value means a 
steeper/gentler quintessential tail. Thus, for a given value of $\varphi_F$, 
we require smaller/larger $n$-values for a smaller/larger-$\alpha$ value. 

\section{Suppressed Interactions}
\label{sec:supint}

In general, quintessence models require an extremely flat potential over 
super-Planckian distances. This gives rise to two problems. First, the 
flatness of such a potential can be lifted by sizeable radiative corrections.
Second, because the mass of the quintessence field is extremely small
\mbox{$\sim H_0\sim 10^{-33}\,$eV}, the corresponding wavelength is very large
(Horizon sized), which can give rise to the infamous 5th force problem, that 
amounts to sizable violations of the Equivalence Principle.

However, in the context of $\alpha$-attractors both the above dangers are 
averted. Indeed, as discussed in Ref.~\cite{Linde:2017pwt,Kallosh:2016gqp}, when near the kinetic poles 
(\mbox{$\phi/\mpl \approx \pm\sqrt{6\alpha}$}, equivalently 
\mbox{$|\varphi|/\mpl\gg\sqrt{6\alpha}$}), the inflaton interactions are 
exponentially suppressed and the field becomes ``asymptotically free''. The
same is true for the loop corrections to the potential. We now briefly 
demonstrate this regarding the interactions.

We expect the inflaton to have Planck-suppressed interactions with other fields.
Following Ref.~\cite{Linde:2017pwt,Kallosh:2016gqp}, lets sketch this by considering another scalar 
field $\sigma$ with which the inflaton is coupled as
\begin{equation}
\delta V=\frac12 h\left(\frac{\phi}{\mpl}\right)^q\phi^2\sigma^2,
\end{equation}
where \mbox{$q\geq 0$} and $h={\cal O}(1)$. Then, the strength of the 
interaction is estimated by 
\mbox{${\cal G}=\partial_\varphi^2\partial_\sigma^2\delta V$}.
It is straightforward to find
\begin{equation}
{\cal G}=
\left(\frac{\partial\phi}{\partial\varphi}\right)^2(q+1)(q+2)h
\left(\frac{\phi}{\mpl}\right)^q.
\end{equation}
Now, near the pole (down the quintessential tail) we have 
\mbox{$\phi/\mpl=\sqrt{6\alpha}$}. Using this and in view of 
Eq.~\eqref{eq:canonicalVariables}, we find
\begin{equation}
{\cal G}=
\frac{(q+1)(q+2)\,h\,(6\alpha)^{q/2}}
{\cosh^4\frac{\varphi_F}{\sqrt{6\alpha}\,\mpl}}
\,.
\label{interq}
\end{equation}
Taking \mbox{$q\sim h\sim\alpha\sim 1$} and \mbox{$\varphi_F\gg\mpl$} we obtain
that the strength of the interaction is suppressed as
\begin{equation}
{\cal G}\sim
\exp\left(-\frac{4\varphi_F}{\sqrt{6\alpha}\,\mpl}\right).
\label{Geq}
\end{equation}
It should be noted here that this suppression is not due to assuming a
Planck-suppressed interaction, as can be readily seen by taking \mbox{$q=0$}
in Eq.~\eqref{interq}. 

\begin{figure}[t]
	\centering
	\includegraphics[width=1.0\linewidth]{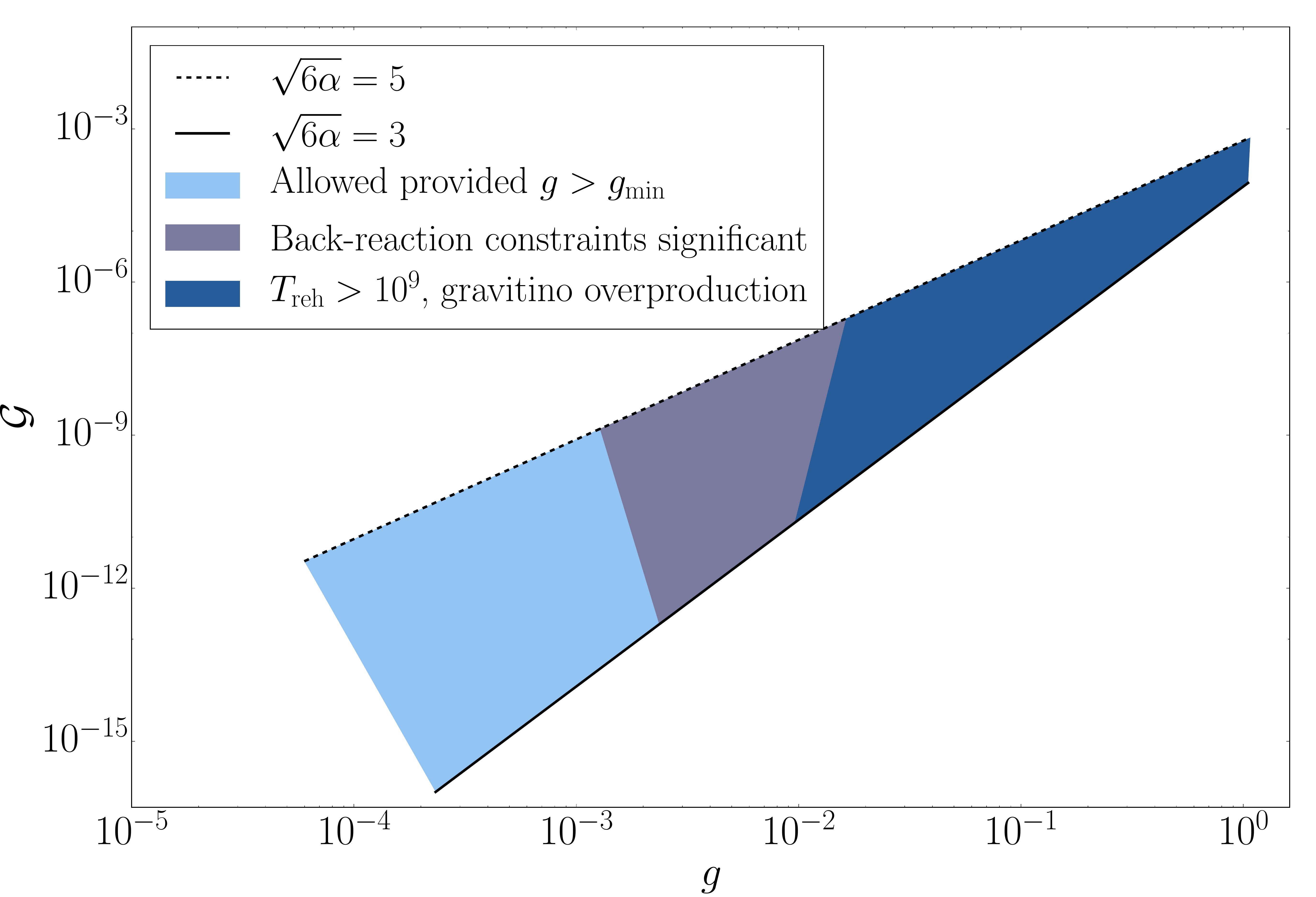}
	\caption{The interaction strength, $\cal G$ for the range of allowed $g$ values and allowed $\alpha$ values between 1.5 and 4.2. The bounds arising from backreaction and gravitino constraints are indicated.}
	\label{fig:G}
\end{figure}

It is straightforward to obtain an estimate of the above value of $\cal G$. 
Indeed, ignoring $\varphi_{\rm IP}$ and using Eq.~\eqref{OIP} we have
\begin{equation}
\varphi_F/\mpl\simeq\sqrt\frac23\left[1-3\ln\left(g/2\pi^{3/2}\right)\right].
\end{equation}
Inserting the above into Eq.~\eqref{Geq}, we obtain
\begin{equation}
{\cal G}\sim e^{-4/3\sqrt\alpha}\left(\frac{g}{2\pi^{3/2}}\right)^{4/\sqrt\alpha}
\end{equation}
Using this we obtain the values shown in Fig~\ref{fig:G}, which demonstrates
that the interaction strength is drastically diminished. 
The above argument can be generalised to non-perturbative 
interactions, which are expected to be of the form 
$\sim\exp(-\beta_i\phi/\mpl){\cal L}_i$, where ${\cal L}_i$ is any 4-dimensional
Lorentz-invariant operator. Considering the interaction strength, we always 
obtain a factor \mbox{$\Big(\frac{\partial\phi}{\partial\varphi}\Big)^2
\sim\exp\left(-\frac{4\varphi_F}{\sqrt{6\alpha}\,\mpl}\right)$}
in the limit \mbox{$\phi/\mpl\rightarrow\sqrt{6\alpha}$}. 
As shown in Fig.~\ref{fig:G}, the 
interaction strength is exponentially suppressed,
which overcomes the 5th force problem. 


In a similar manner, loop corrections are also suppressed so the flatness
of the quintessential tail is safely protected from radiative corrections
\cite{Linde:2017pwt,Kallosh:2016gqp}.

\section{Conclusions}

We have investigated a model of quintessential inflation in the context of 
$\alpha$-attractors in supergravity. We considered a simple exponential
potential \mbox{$V(\phi)=V_0 e^{-\kappa\phi/m_P}$} (c.f. Eq.~\eqref{LV0}) and 
the standard $\alpha$-attractors kinetic term, 
which features two poles at \mbox{$\phi=\pm\sqrt{6\alpha}\,\mpl$} 
(c.f. Eq.~\eqref{Lkin}). Switching to a canonically normalised inflaton, the 
scalar potential gets ``stretched'' as the poles are transposed to infinity 
\cite{Kallosh:2013yoa,Kallosh:2013tua,Ferrara:2013rsa,Ferrara:2013kca}, thereby generating the inflationary plateau and the quintessential
tail. After inflation, the field becomes kinetically dominated as it ``jumps
off the cliff'' of the inflationary plateau. A period of kination ensues. This 
necessarily ends when the Universe becomes dominated by radiation and the hot big 
bang begins. This radiation is generated through the mechanism of instant 
preheating. For this we assume that the inflaton $\phi$ is coupled with some 
other scalar field $\chi$ such that, after the end of inflation when the 
inflaton's variation peaks, the effective mass of the $\chi$-particles is
varying non-adiabatically. This adiabaticity breaking results in particle 
production of $\chi$-particles, which soon decay into a newly formed radiation 
bath. The strength of the interaction between $\phi$ and $\chi$ is parametrised 
by the coupling $g$ (c.f. Eq.~\eqref{IPL}).

\begin{figure}[t]
	\centering
	\includegraphics[width=1\linewidth]{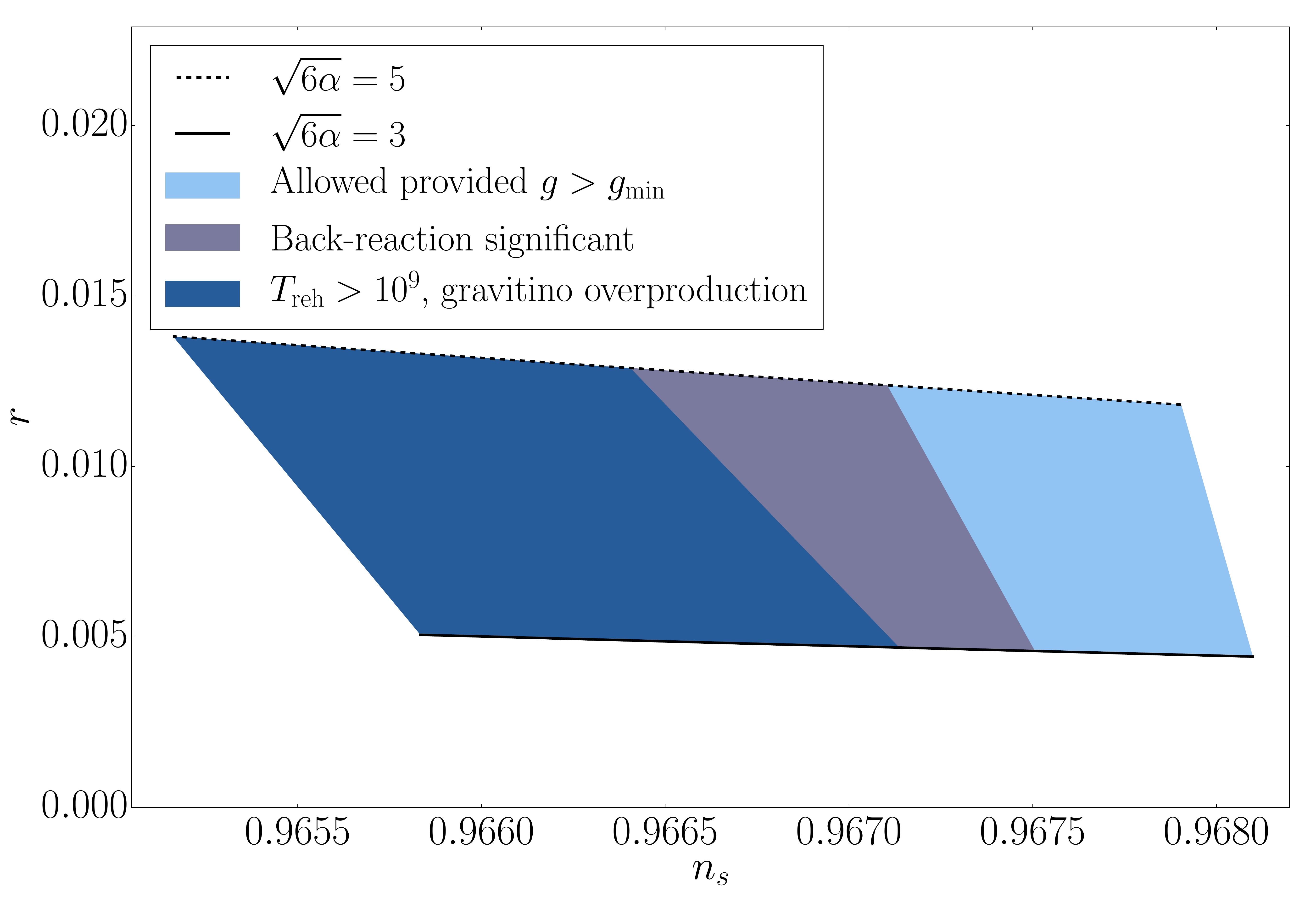}
	\caption{$n_s$ and $r$ values for the range of $n$ values indicated in Eq.\eqref{eq:n_large_param_space}, for allowed $\alpha$ values between 1.5 and 4.2. The bounds arising from backreaction and gravitino constraints are indicated.}
	\label{fig:ns_r}
\end{figure}

\begin{figure}[t]
	\centering
	\includegraphics[width=1\linewidth]{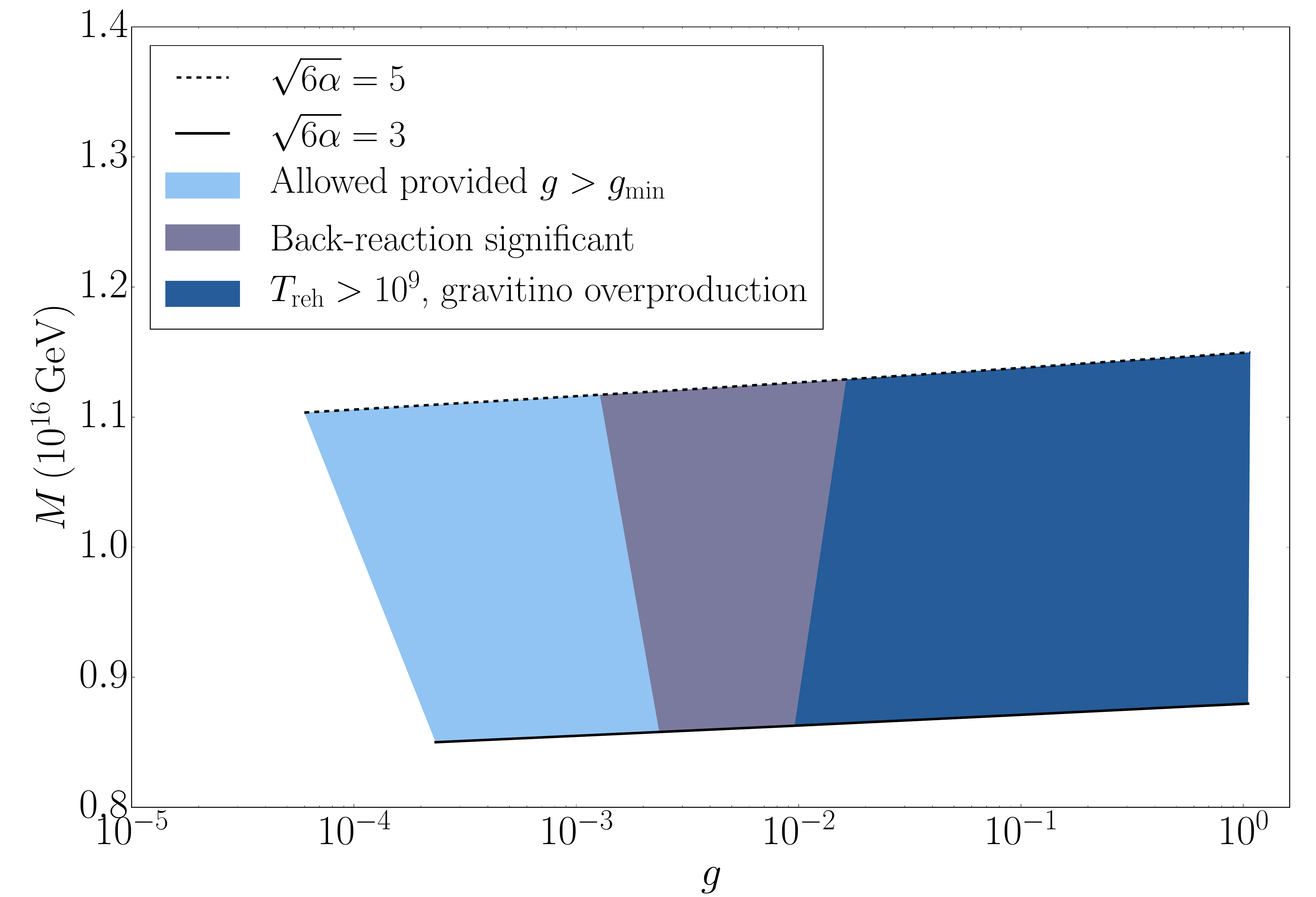}
	\caption{$M$ values for the range of $g$ values and allowed $\alpha$ values between 1.5 and 4.2. The bounds arising from backreaction and gravitino constraints are indicated.}
	\label{fig:M}
\end{figure}

\begin{figure}[t]
	\centering
	\includegraphics[width=1\linewidth]{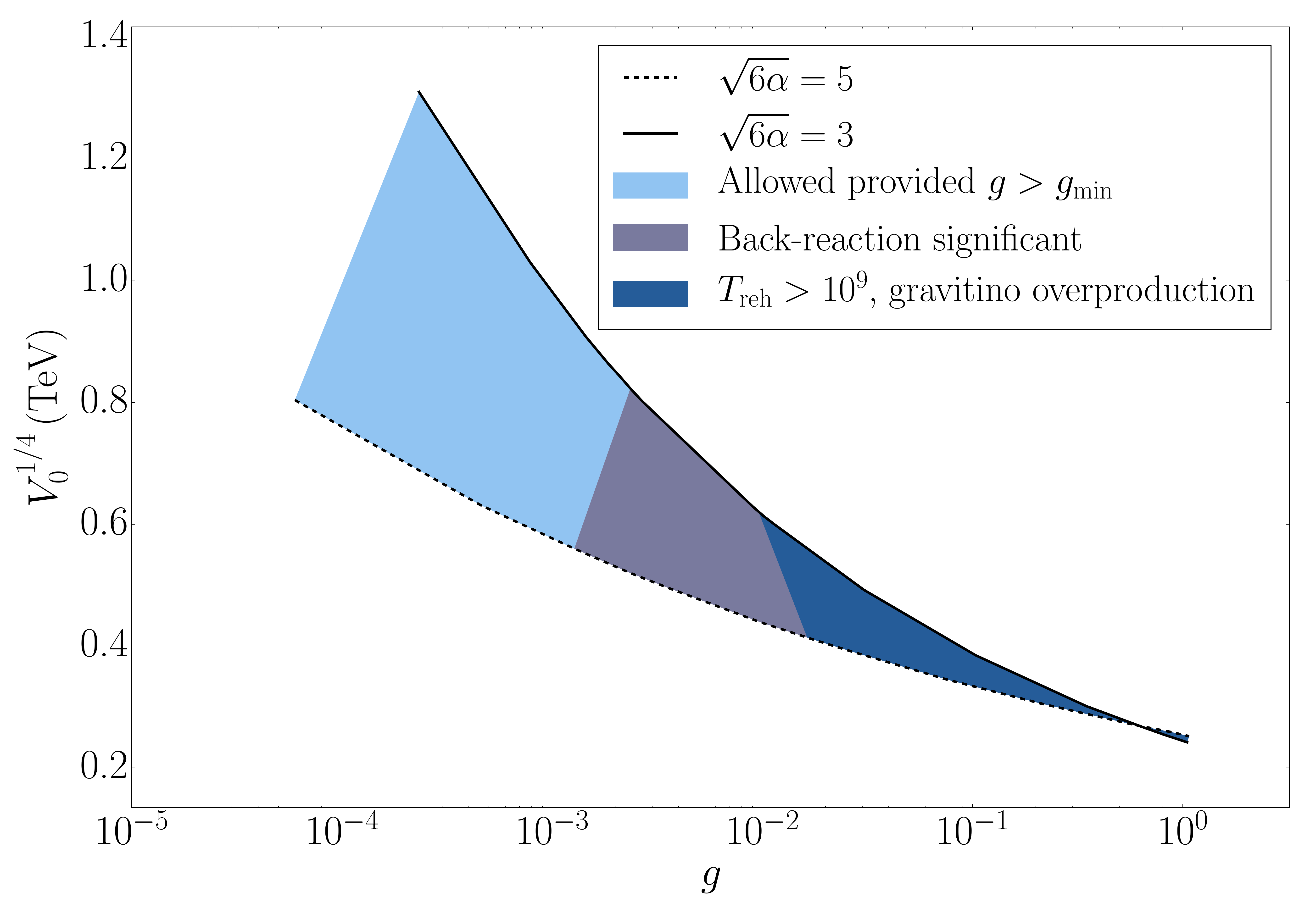}
	\caption{$V_0^{1/4}$ values for the range of $g$ values and allowed $\alpha$ values between 1.5 and 4.2. The bounds arising from backreaction and gravitino constraints are indicated.}
	\label{fig:V0}
\end{figure}

We have investigated the parameter space available, when the observational
constraints on the abundance and barotropic (equation of state) parameter of
dark energy are considered. We were conservative in avoiding a 
super-Planckian inflaton field $\phi$, even though the suppression of loop 
corrections and interactions of the inflaton near the poles in 
$\alpha$-attractors \cite{Linde:2017pwt,Kallosh:2016gqp} would mean that, even if the inflaton were 
super-Planckian, the flatness of the quintessential runaway potential would be 
preserved and there would not be a fifth-force problem. Moreover, we have taken into account backreaction constraints, 
which threaten to shut down $\chi$-particle production and gravitino 
constraints on the reheating temperature.

When all the constraints are applied
we find that our model is successful for natural values of the model parameters.
In particular, for the coupling we find \mbox{$g\sim 10^{-4}-10^{-2}$}, while we 
also have \mbox{$V_0^{1/4}\sim 1\,$TeV}, which is the electroweak energy scale (Fig.~\ref{fig:V0}). 
The inflationary scale is \mbox{$M\simeq 10^{16}\,$GeV}, which is at the energy 
scale of grand unification (Fig.~\ref{fig:M}). 
For the slope of the exponential potential we find 
\mbox{$\kappa\simeq 24-40$} (Fig~\ref{fig:kappa}), i.e. 
\mbox{$\kappa\sim 0.1\mpl/M$}, meaning that in the potential the inflaton is 
suppressed by the scale \mbox{$\sim 10^{17}\,$GeV} (string scale?).

We also find that the cosmological scales exit the horizon
about \mbox{$N_*\simeq 62-63$} e-folds before the end of inflation (Fig.~\ref{fig:Nstar}) and that the
reheating temperature is \mbox{$T_{\rm reh}\sim 10^5-10^8\,$GeV} (Fig.~\ref{fig:Treh}), which satisfies 
gravitino constraints as required. 
For the inflationary observables we obtain 
the values \mbox{$n_s=0.968$} for the spectral index and 
\mbox{$n_s'=-(5-6)\times 10^{-4}$} for its running.  
For the tensor to scalar ratio we obtain
\mbox{$r\simeq 0.004 - 0.012$}, which may well be observable (Fig.~\ref{fig:ns_r}). These values are within the 1-$\sigma$ contour of the Planck results \cite{Ade:2015lrj}.

The $\alpha$-attractors setup may also be realised without relying on 
supergravity \cite{Kallosh:2013yoa,Kallosh:2013tua,Ferrara:2013rsa,Ferrara:2013kca}.
In this case, the gravitino constraints may not be necessary. Also, backreaction
effects can be dispensed with when the $\chi$-particles decay rapidly into 
radiation, such that they don't backreact and close the resonance. If we remove 
these constraints, our parameter space is substantially enlarged. In particular,
$g$ can approach unity, while 
$N_*$ can be as low as
\mbox{$N_*\simeq 59$} and the reheating temperature can be as large as
\mbox{$T_{\rm reh}\sim 10^{11}\,$GeV}. 
Regarding the inflationary observables, the spectral index can become as low as
\mbox{$n_s=0.966$}, but $r$ is not changed much.

\begin{figure}[t]
	\centering
	\includegraphics[width=1\linewidth]{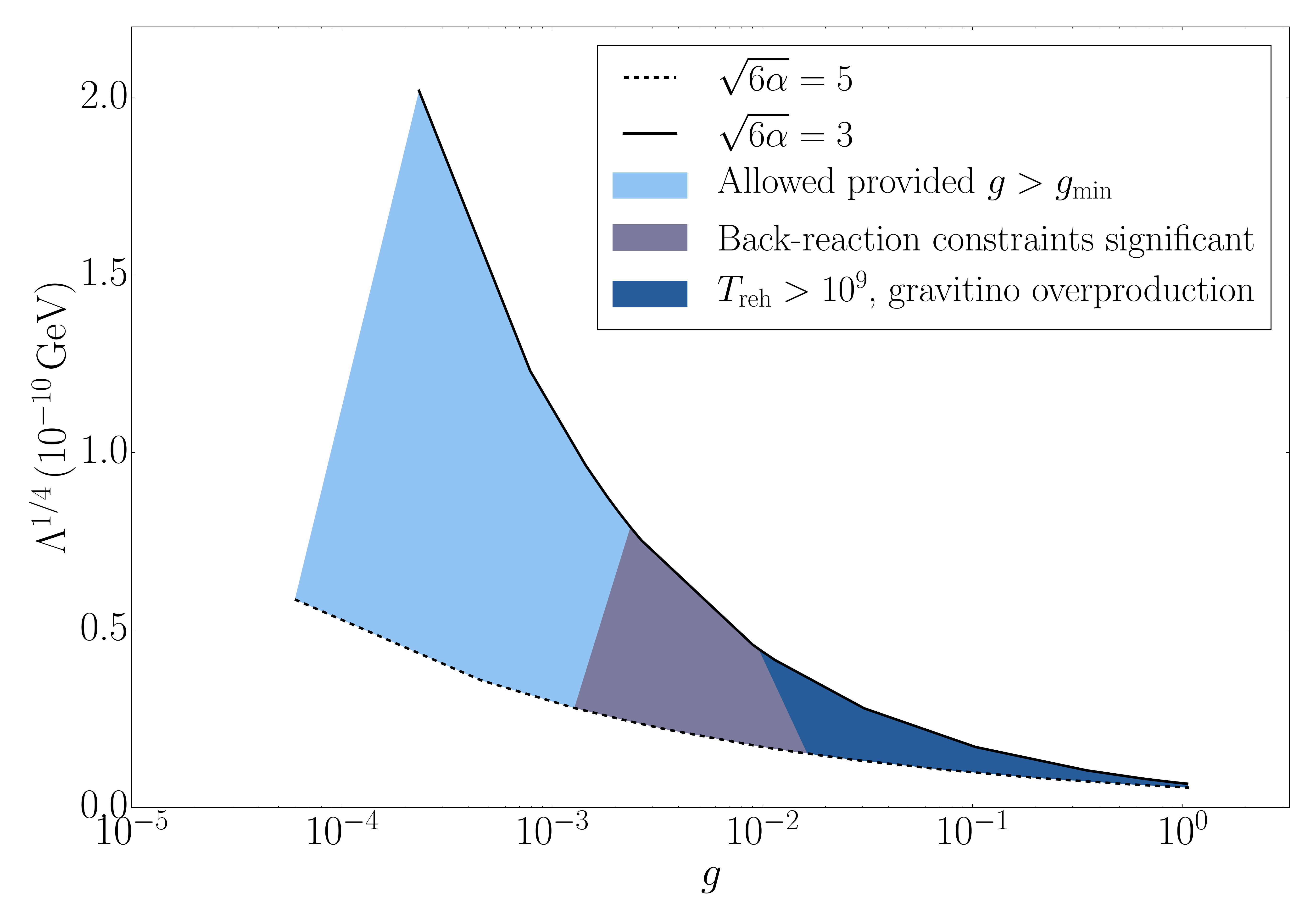}
	\caption{$\Lambda^{1/4}$ values for the range of $g$ values and allowed $\alpha$ values between 1.5 and 4.2. The bounds arising from backreaction and gravitino constraints are indicated.}
	\label{fig:Lambda}
\end{figure}

The required cosmological constant is
\mbox{$\Lambda^{1/4}\sim 10^{-10}\,$GeV} (Fig.~\ref{fig:Lambda}), 
which is somewhat larger that the value $\sim 10^{-3}\,$eV required in 
$\Lambda$CDM, but the improvement is not much. One may worry that, if one is 
prepared to accept the scale $0.1\,$eV, why not stay with $\Lambda$CDM in the 
first place. The answer is two-fold. Firstly, in contrast to $\Lambda$CDM,
our required value for $\Lambda$ is not imposed ad~hoc to satisfy the 
observations. Instead, it is generated by the requirement that the vacuum 
energy asymptotes to zero (cf. Eq.~\eqref{Lambda}). In other words, the 
(unknown) mechanism which demands
zero vacuum density is the one which imposes our value of $\Lambda$. The
second reason has to do with the future horizon problem in string theories \cite{Hellerman:2001yi,Fischler:2001yj,Banks:2000fe,Banks:2001yp}.
In a nutshell, in $\Lambda$CDM, there is a future event horizon, which 
makes the asymptotic future states not well defined because they are not
causally connected. As a result, the formulation of the S-matrix is 
problematic \cite{Dyson:2002pf,Goheer:2002vf}. Our model may overcome this problem as follows:
Since the eventual value of the vacuum density is zero, this means that the
size of the future event horizon increases to infinity. Thus, future states 
are well defined and the future horizon problem is overcome. Finally, it is 
important to point out that our model considers a varying barotropic parameter 
of dark energy, which will be tested in the near future.

In summary, we have shown that our model of quintessential inflation with
$\alpha$-attractors, first introduced in Ref.~\cite{Dimopoulos:2017zvq}, works 
well with instant preheating, improving the robustness of the model. 


\noindent
{\em Note:} After our paper originally appeared in the arXiv, Ref.~\cite{Akrami:2017cir} came out, which considers quintessential inflation with $\alpha$-attractors beyond the single-field exponential potential.

\paragraph{Acknowledgements}
KD is supported (in part) by the Lancaster-Manchester-Sheffield Consortium for Fundamental Physics under STFC grant: ST/L000520/1. CO is supported by the FST of Lancaster University and would like to thank the Physics Department at The Johns Hopkins University for their hospitality whilst this paper was completed.

\bibliography{IP_in_Quint_Refs2}

\appendix

\section*{Appendix: The Range of \boldmath $\alpha$}\label{sec:App}

In this Appendix we calculate the appropriate range for $\alpha$, upon which 
our results are based. To do this, we first investigate exponential 
quintessence.

We consider single field quintessence with a canonical scalar field $\varphi$ 
and a scalar potential $V(\varphi)$, which drives the currently observed 
accelerated expansion. This simple model assumes only a minimal coupling 
between gravity and $\varphi$, and is thus described by the action
\begin{equation}\label{eq:app_action}
S = \int \mathrm{d}^4x\sqrt{-g}\mathcal{L} + S_m(g_{\mu\nu};\Psi_n) \,,
\end{equation}
where the scalar tensor Lagrangian density is
\begin{equation}\label{eq:app_langs}
\mathcal{L} = \frac{1}{2}\mpl^2R - \frac{1}{2}(\partial\varphi)^2 - V(\varphi) \,,
\end{equation}
and $S_m$ is the action for any matter fields present, $\Psi_n$, coupled to 
gravity. 

To explore the dynamics of any late universe quintessence model we assume a 
$w$CDM cosmology.  We have an FRW metric and assume the effects of $\rho_r$ 
are negligible and that $\rho_{\Lambda}=0$. As such, the content of the Universe 
is modelled as two perfect fluid components; our scalar field $\varphi$, and a 
non-relativistic background matter fluid, denoted by subscript `$m$', with 
equations of state \mbox{$p_i = w_i\rho_i, i \in \{\varphi,m\}$} where 
\mbox{$w_{\varphi} = w_{\varphi}(t)$} and \mbox{$p_m = 0 \Rightarrow w_m = 0$}.  
Ignoring perturbations and any spatial curvature, 
because \mbox{$\Omega_{\mathrm{K}} \simeq 0$} \cite{Ade2016}, we have
\begin{equation}\label{eq:app_metric}
g_{\mu\nu}\mathrm{d}x^{\mu}\mathrm{d}x^{\nu} = -\mathrm{d}t^2 + a^2(t)\delta_{ij}\mathrm{d}x^i\mathrm{d}x^j \,,
\end{equation}
\begin{equation}\label{eq:app_rhos_ps}
\rho_{\varphi} = \frac{1}{2}\dot{\varphi}^2 + V(\varphi), 	\quad p_{\varphi} = \frac{1}{2}\dot{\varphi}^2 - V(\varphi)\,,	
\end{equation}
\begin{equation}\label{eq:app_ws}
w_{\varphi} = \frac{p_{\varphi}}{\rho_{\varphi}} = \frac{\frac{1}{2}\dot{\varphi}^2 - V(\varphi)}{\frac{1}{2}\dot{\varphi}^2 + V(\varphi)}, 	\qquad w = \frac{\Sigma_i p_i}{\Sigma_i \rho_i} = \frac{p_{\varphi}}{\rho} \,,
\end{equation}
where $\rho = \Sigma_i \rho_i = \rho_m + \rho_{\varphi}$. The evolution equations are
\begin{equation}\label{eq:app_Fried_evo}
-2\dot{H}\mpl^2 = \dot{\varphi}^2 + \rho_m \,,
\end{equation}
\begin{equation}\label{eq:app_rho_evo}
\dot{\rho}_m = -3H\rho_m \,,
\end{equation}
\begin{equation}\label{eq:app_KG}
\frac{\delta S}{\delta \varphi} = \ddot{\varphi} + 3H\dot{\varphi} + V'(\varphi) = 0 \,,
\end{equation}
conditional on the Friedman equation
\begin{equation}\label{eq:app_Fried}
3\mpl^2H^2 = \frac{1}{2}\dot{\varphi}^2 + V(\varphi) + \rho_m \,.
\end{equation}

\pagebreak
Specific quintessence models are distinguished by considering suitable forms of $V(\varphi)$ (typically of 
runaway type) which are flat enough to lead to the current accelerated 
expansion at late times.  In quintessential inflation models, this region is 
called the quintessential tail. Quintessential inflation considers so-called 
``thawing'' quintessence, where, until recently, $\varphi$ was frozen, but near 
the present it unfreezes and begins to evolve into a possible slow-roll regime 
as it dominates the Universe and drives the current accelerated expansion.


Here we consider the specific case of exponential quintessence where
\begin{equation}\label{eq:app_V}
V(\varphi) = V_Q\,\mathrm{exp}(-\lambda\varphi/\mpl) \,,
\end{equation}
where $V_Q$ is a constant density scale and $\lambda$ is a constant parameter.

The Klein-Gordon equation, Eq. \eqref{eq:app_KG} admits two attractor solutions 
which depend on the eventual dominance vs. sub-dominance of $\varphi$ with 
regard to the background matter:
\begin{eqnarray}\label{eq:app_attr_sol_1}
\mathrm{Dominant} \,\, \Big(\mathrm{for} \, \lambda<\sqrt{3(1+w_b)}\Big): 
\nonumber \\ 
\quad V = \frac{2(6-\lambda^2)}{\lambda^4}\Big(\frac{\mpl}{t}\Big)^2\;\&\;
\rho_{\mathrm{kin}} 
= \frac{2}{\lambda^2}\Big(\frac{\mpl}{t}\Big)^2 \nonumber\\
\Rightarrow\;\rho_{\varphi} = \frac{12}{\lambda^4}\Big(\frac{\mpl}{t}\Big)^2 \,. 
\label{eq:app_attr_sol_1_rho} 
\end{eqnarray}
\begin{eqnarray}\label{eq:app_attr_sol_2}
\mathrm{Subdominant} \,\, \Big(\mathrm{for} \, \lambda>\sqrt{3(1+w_b)}\Big): 
\nonumber \\ 
\quad V=\frac{2}{\lambda^2}\Big(\frac{1-w_b}{1+w_b}\Big)
\Big(\frac{\mpl}{t}\Big)^2\;\&\;
\rho_{\mathrm{kin}} = \frac{2}{\lambda^2}\Big(\frac{\mpl}{t}\Big)^2\nonumber\\
\Rightarrow\;\rho_{\varphi}=\frac{4}{\lambda^2(1+w_b)}\Big(\frac{\mpl}{t}\Big)^2
\,,
\label{eq:app_attr_sol_2_rho}
\end{eqnarray}
where \mbox{$\rho_{\mathrm{kin}}\equiv\frac{1}{2}\dot{\varphi}^2$} and $w_b$ is
the barotropic parameter of the background; being \mbox{$w_b=0$} for matter.

The solutions differ with regard to the evolution of $\rho_{\varphi}$ in 
comparison to that of $\rho_m$, which is fixed at $\rho_m \propto a^{-3}$.  For 
dominant quintessence Eq.~\eqref{eq:app_attr_sol_1_rho}, 
\mbox{$\rho_{\varphi}\propto a^{-\lambda^2}$}, and for subdominant quintessence 
Eq.~\eqref{eq:app_attr_sol_2_rho}, \mbox{$\rho_\varphi\propto a^{-3}$}.  The 
subdominant quintessence attractor solution is called a scaling solution 
because $\rho_{\varphi}/\rho_m$ stays constant. The value of $\lambda$ determines 
both the slope of the quintessential tail, and which attractor solution the 
field eventually follows. The value of $\lambda=\sqrt{3}$ (since \mbox{$w_b=0$})
represents the boundary between the two attractor solutions, i.e. as $\lambda$ 
increases toward $\sqrt{3}$, the evolution of $\rho_{\varphi}$ increasingly moves 
towards that of $\rho_{\varphi}\propto a^{-3}$.

Copeland et al \cite{Copeland:1997et} used a phase-plane analysis and found that, for 
$\lambda<\sqrt{3}$ the dominant quintessence attractor solution 
(Eq.~\eqref{eq:app_attr_sol_1}) is a stable node. For 
\mbox{$\sqrt{3}<\lambda<\sqrt{6}$} and \mbox{$\lambda>\sqrt{6}$} the subdominant quintessence
attractor solution (Eq.~\eqref{eq:app_attr_sol_2}) is a stable node/spiral and 
a stable spiral respectively. After unfreezing, the field briefly oscillates 
about the attractor before settling on the attractor solution.  For 
\mbox{$\lambda<\sqrt{3}$}, it is easy to show that \mbox{$w=-1+\lambda^2/3$} on 
the attractor \cite{Dimopoulos:2017zvq}. This means that $\lambda<\sqrt{2}$ results in 
$w<-1/3$, which leads to eternal accelerated expansion.  For 
\mbox{$\sqrt{2}\lesssim\lambda<\sqrt{3}$} 
(\mbox{$\sqrt{3}\lesssim\lambda<2\sqrt{6}$}), the brief oscillation of the 
field about the dominant quintessence (subdominant quintessence) attractor, may 
result in a bout of transient accelerated expansion \cite{Dimopoulos2003,Franca2002,Cline2001,Kolda2001,Kehagias2004}. 

We numerically explore the cosmological dynamics of this single field 
quintessence model, to a confirmed accuracy of $10^{-4}$(4 d.p) for all 
cosmological parameters.  We use the latest Planck observations to constrain 
the range of $\lambda$ for which any current eternal or transient accelerated 
expansion is present.  The latest Planck observations \cite{Ade2016} suggest that the density parameter of dark energy is 
\mbox{$\Omega_{\Lambda}=1-\Omega_{\rm K}-\Omega_m$}, where 
\mbox{$\Omega_{\rm K}=0.000\pm 0.005$}, and \mbox{$\Omega_m=0.308\pm 0.012$}. 
This results in \mbox{$\Omega_{\Lambda}=0.692\pm0.017$}.

As we have a time-varying $w_{\varphi}$, we model a Taylor expansion of 
$w_{\varphi}$ to first order 
\begin{equation}\label{eq:app_w_taylor}
w_\varphi = w_{\mathrm{DE}} + \Big(1 - \frac{a}{a_0}\Big)w_a \,,
\end{equation}
where \mbox{$w_{a}=-(\mathrm{d}w_{\varphi}/\mathrm{d}a)_0 = -\dot{w}_{\mathrm{DE}}$},
the subscript `0' denotes values today, when \mbox{$a=a_0$} and 
\mbox{$w_\varphi(a_0)=w_{\rm{DE}}$}. We  use the Planck bounds \cite{Ade2016} of 
\mbox{$w_{\rm{DE}} = -1.023^{+0.091}_{-0.096}$} at 
\mbox{2-$\sigma$} in our constraint on possible ranges of values for $\lambda$. 
This 
translates to \mbox{$w_0 = -0.7112\pm0.0821$}, where $w_0$ is the barotropic 
parameter of the Universe at present, \mbox{$w_0 = (p_{\varphi}/\rho)_0$} 
(cf. Eq.~\eqref{eq:app_ws}).

Demanding that our model satisfies these observational requirements, the 
Universe today has to lie within the range \mbox{$(\rho_{\varphi}/\rho_{m})_0 = 
\Omega_\Lambda/\Omega_m=2.2523\pm0.1429$}, within which we can investigate any 
current eternal or transient accelerated expansion found. We start with the 
frozen field, where \mbox{$\dot{\varphi}_F=0$} and 
\mbox{$(\rho_{\varphi}/\rho_m)_F\ll 1$}, where the subscript `$F$' denotes frozen values.

Only a change in the value of $\lambda$ affects the evolution of our model once 
the field is unfrozen.  A relative decrease (increase) in the value of $\rho_{\varphi}^F = V(\varphi_F)$, 
for a given $\rho_m^F$, only increases (decreases) the evolution 
time of the model until $a_0$ today, i.e. the model is extended backwards  
(forwards) to an earlier (later) time when $\varphi$ is frozen. Similarly,
any change in the value of $\varphi_F$ can be expressed as a change in 
$V_Q$, and so for a given value of $\lambda$, also has no effect on the 
dynamics. Conversely, since $\Omega_\Lambda$ is fixed by the observations, 
changes in $\varphi_F$ without a change in $V_Q$ 
must instead be 
accompanied with corresponding changes in $\lambda$, such that the contribution 
of quintessence to the density budget at present remains fixed . 

\pagebreak
\subsection*{Transient Accelerated Expansion}

For brief periods of transient accelerated expansion with $w<-1/3$, we find a 
range of numerically valid $\lambda$ values bridging the dominant and 
subdominant quintessence regimes
\begin{equation}
\sqrt{2}\lesssim \lambda < \sqrt{3.38}
\end{equation}

\begin{figure}
	\centering
	\includegraphics[width=1\linewidth]{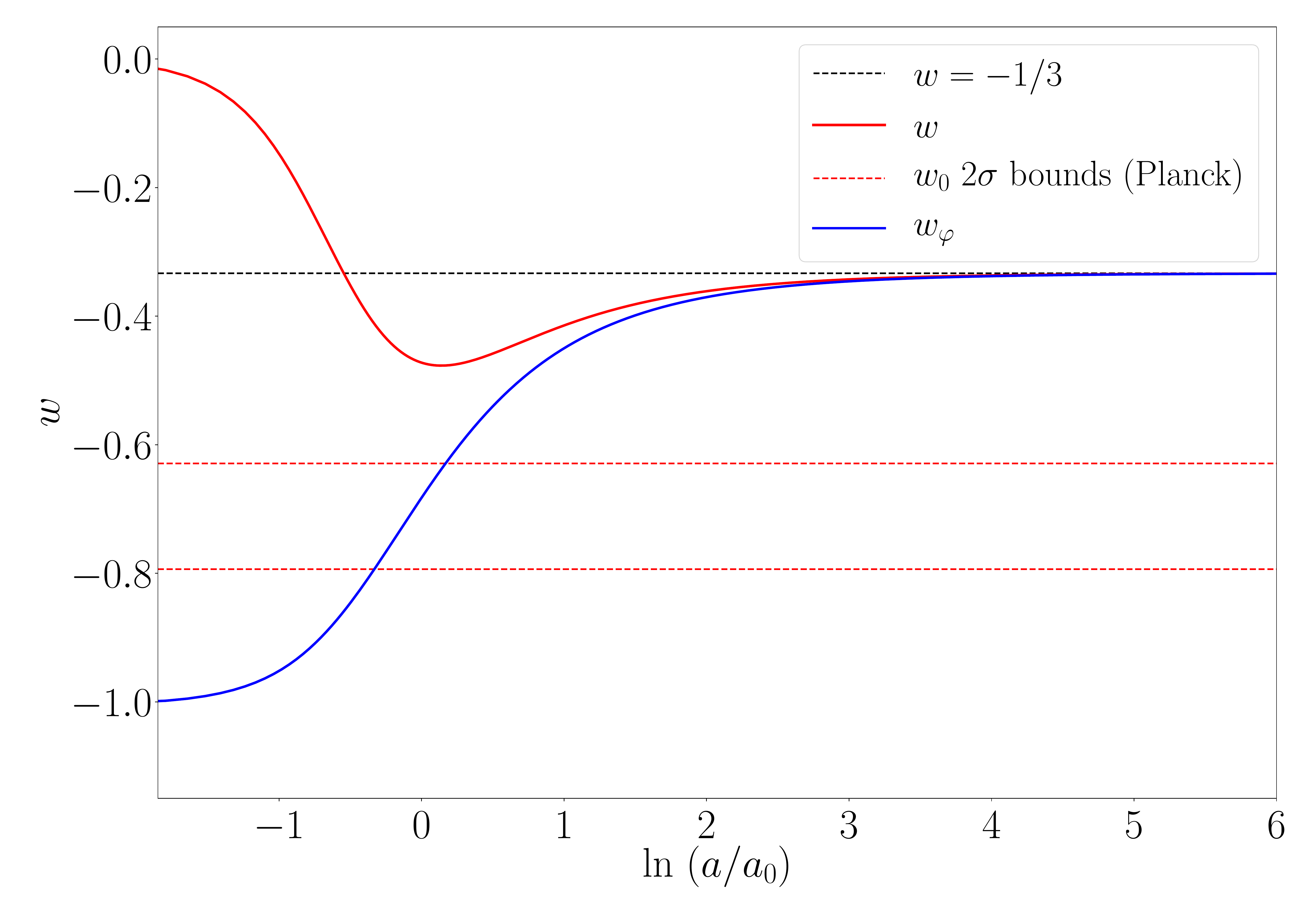}	
\caption{Transient accelerated expansion for $\lambda=\sqrt{2}$. We find 
$w<-1/3$, but the minimum value of $w$ is well outside of the Planck bounds.}
	\label{fig:app_transient}
\end{figure}

However, the values for $w$ that we find in this scenario are incompatible with 
the Planck constraints for  the entire range of $\lambda$ values above. As the 
minimum value of $w$ reached during any period of evolution increases with 
increasing $\lambda$, we only need to look at $\lambda = \sqrt{2}$ to 
illustrate our findings. This is shown in Fig.~\ref{fig:app_transient}, where 
we are using $\lambda = \sqrt{2}$.  It can be clearly seen that the minimum 
value of $w$ is not nearly small enough to match the Planck observational 
bounds, and so all higher values of $\lambda$ are also ruled out.%
\footnote{Figure~\ref{fig:app_transient} also highlights the validity of 
\mbox{$w=-1+\lambda^2/3$} requiring $\lambda<\sqrt{2}$ for eternal accelerated 
expansion, as we can clearly see \mbox{$w=-1/3$} 
in the attractor limit where $\lambda=\sqrt{2}$. We can also see $w_\varphi$ moving toward the 
same value because we are in the dominant quintessence regime.}

\subsection*{Eternal Accelerated Expansion}

\begin{figure}[t]
	\centering
	\includegraphics[width=1\linewidth]{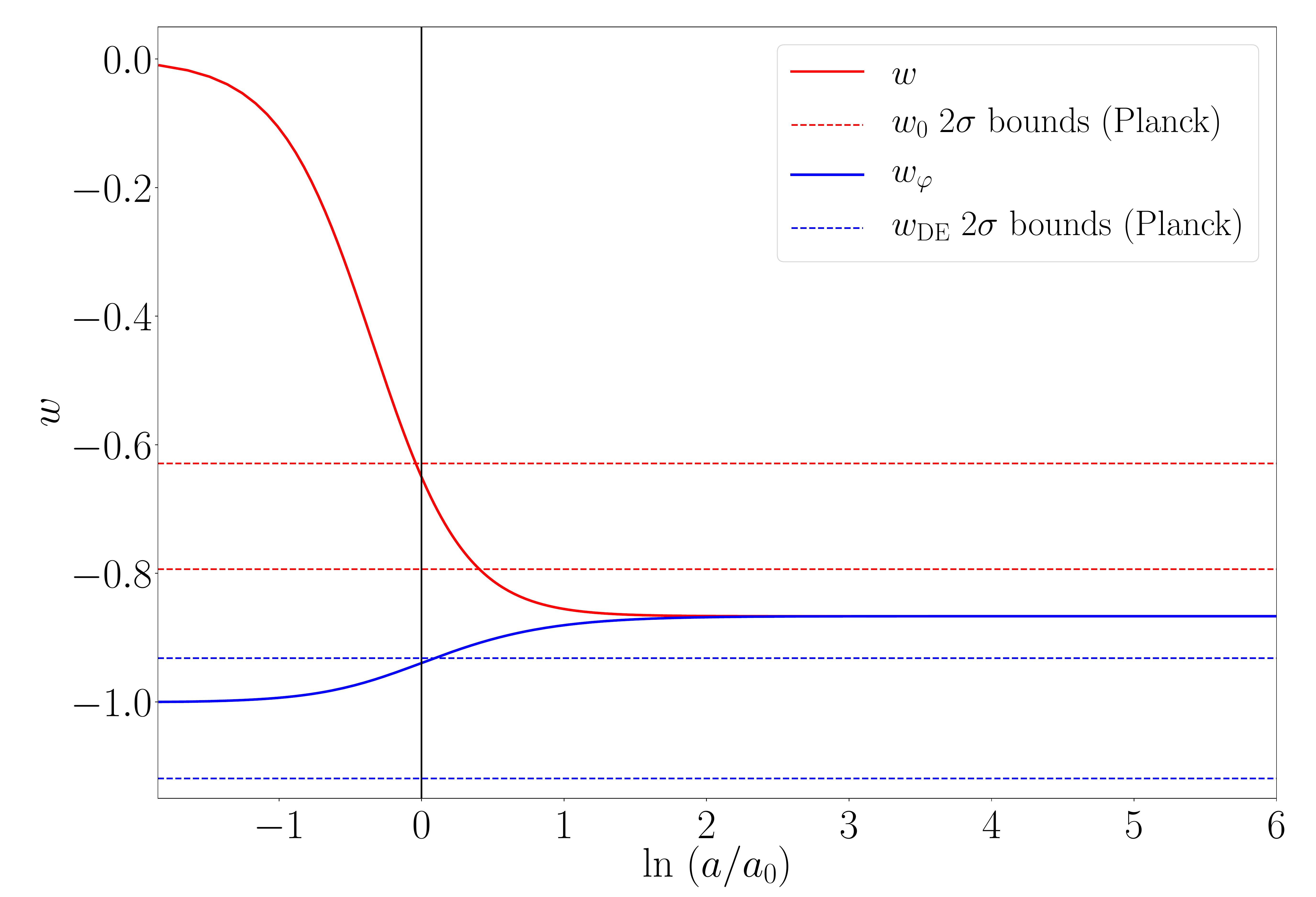}
\caption{Eternal accelerated expansion for $\lambda=\sqrt{0.4}$. The vertical 
line \mbox{$\mathrm{ln}(a/a_0)=0$} indicates present day values of $w$ and 
$w_{\varphi}$ ($w_0$ and $w_{\rm DE}$ respectively), which also fall within the 
required Planck bounds for $w$ and 
$w_{\varphi}$ today. The scalar field has unfrozen, but is yet to settle on the 
attractor solution.}
	\label{fig:app_eternal}
\end{figure}

We know theoretically that $w<-1/3$ for $\lambda<\sqrt{2}$.  When applying the 
Planck constraints we find that the cosmologically viable range is reduced to 
of \mbox{$\lambda<\sqrt{0.46}$}.  We find that, in all cases, the scalar field 
at present has unfrozen but is yet to settle on the attractor solution.  This 
is illustrated in Fig.~\ref{fig:app_eternal} for \mbox{$\lambda=\sqrt{0.4}$}, 
where it can be clearly seen the field has yet to evolve to its attractor 
solution. It can also be clearly seen that the present day values at 
\mbox{$\mathrm{ln}(a/a_0) = 0$} are within the Planck bounds.

\begin{center}
	\begin{figure}[h]
		\includegraphics[scale=0.3]{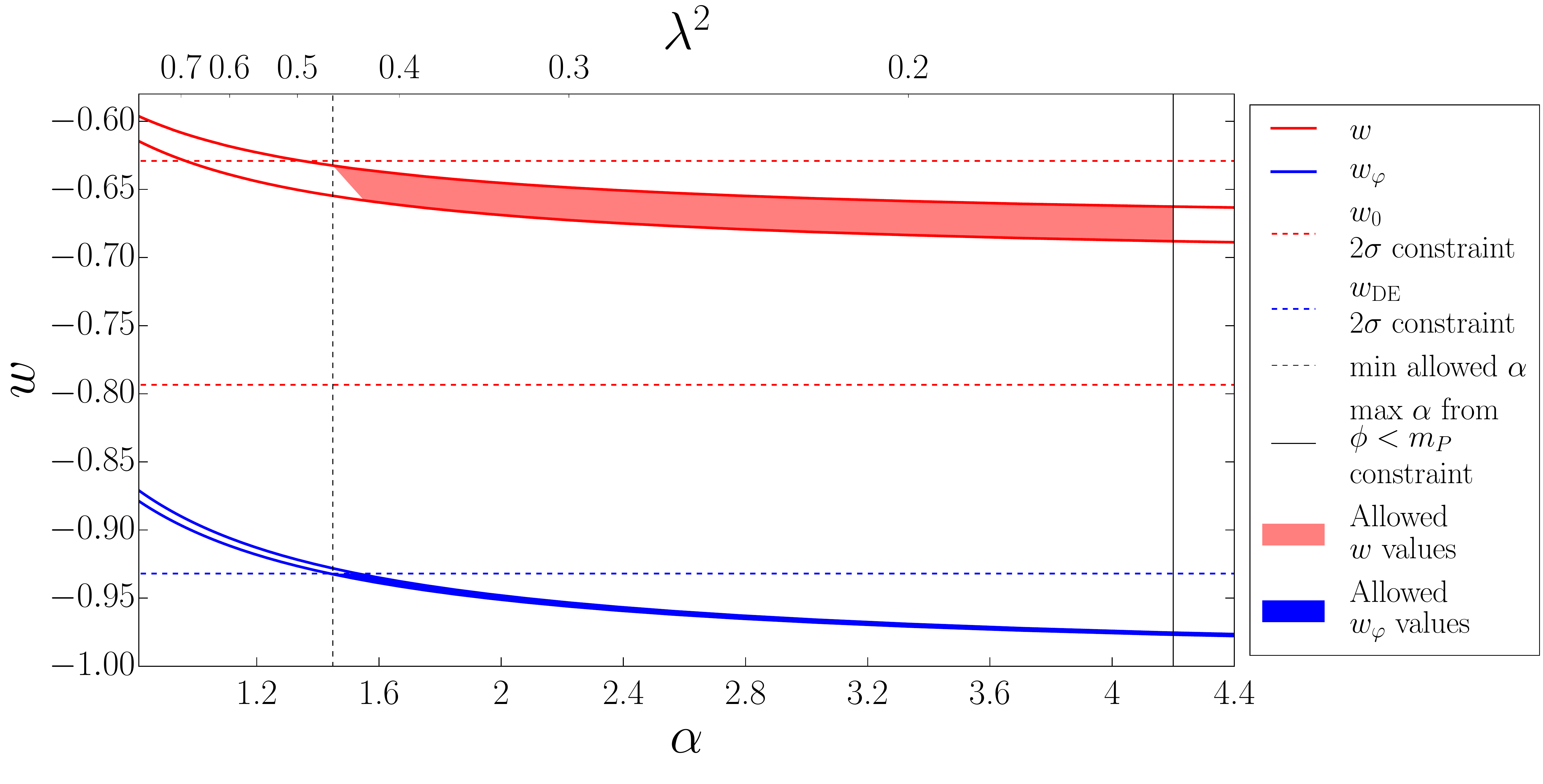}
		\caption{Possible range of values for $\alpha$ and $\lambda^2$, from the 
			Planck constraints on $w$. It is shown that the 2-$\sigma$ upper bound on
			$w_{\rm DE}$ is satisfied only for \mbox{$\lambda^2<0.46$} or equivalently
			\mbox{$\alpha=2/3\lambda^2>1.45$}. The allowed ranges of $w$ and $w_{\varphi}$
			reflect the observed range in $\Omega_\Lambda/\Omega_m$. Values in the text are quoted to 2 s.f.}
			\label{fig:app_amendola}
	\end{figure}
\end{center}

As illustrated in Fig.~\ref{fig:app_amendola}, we find that it is the bound 
for \mbox{$w_{\rm{DE}}=-1.023_{-0.096}^{+0.091}$} that constrains our possible range 
of values to \mbox{$\lambda < \sqrt{0.46}$}. This can also be seen in 
Fig.~\ref{fig:app_eternal}, where the value of $w_{\rm{DE}}$ is closer to the 
upper Planck bound for $w_{\mathrm{DE}}$ compared to the value for $w_0$, which is 
further within the upper Planck bound for $w_0$. When increasing $\lambda$, we 
find that $w_{\mathrm{DE}}$ exits the upper Planck bound for $w_{\mathrm{DE}}$ 
before $w_0$ exits the upper 
Planck bound for $w_0$. If we ignore this constraint and just demand that 
\mbox{$w_0=-0.7112 \pm 0.0821$} today, then our range of possible values for 
$\lambda$ extends 
to \mbox{$\lambda < \sqrt{0.68}$}. 

\begin{figure}[]
	\centering
	\includegraphics[width=0.99\linewidth]{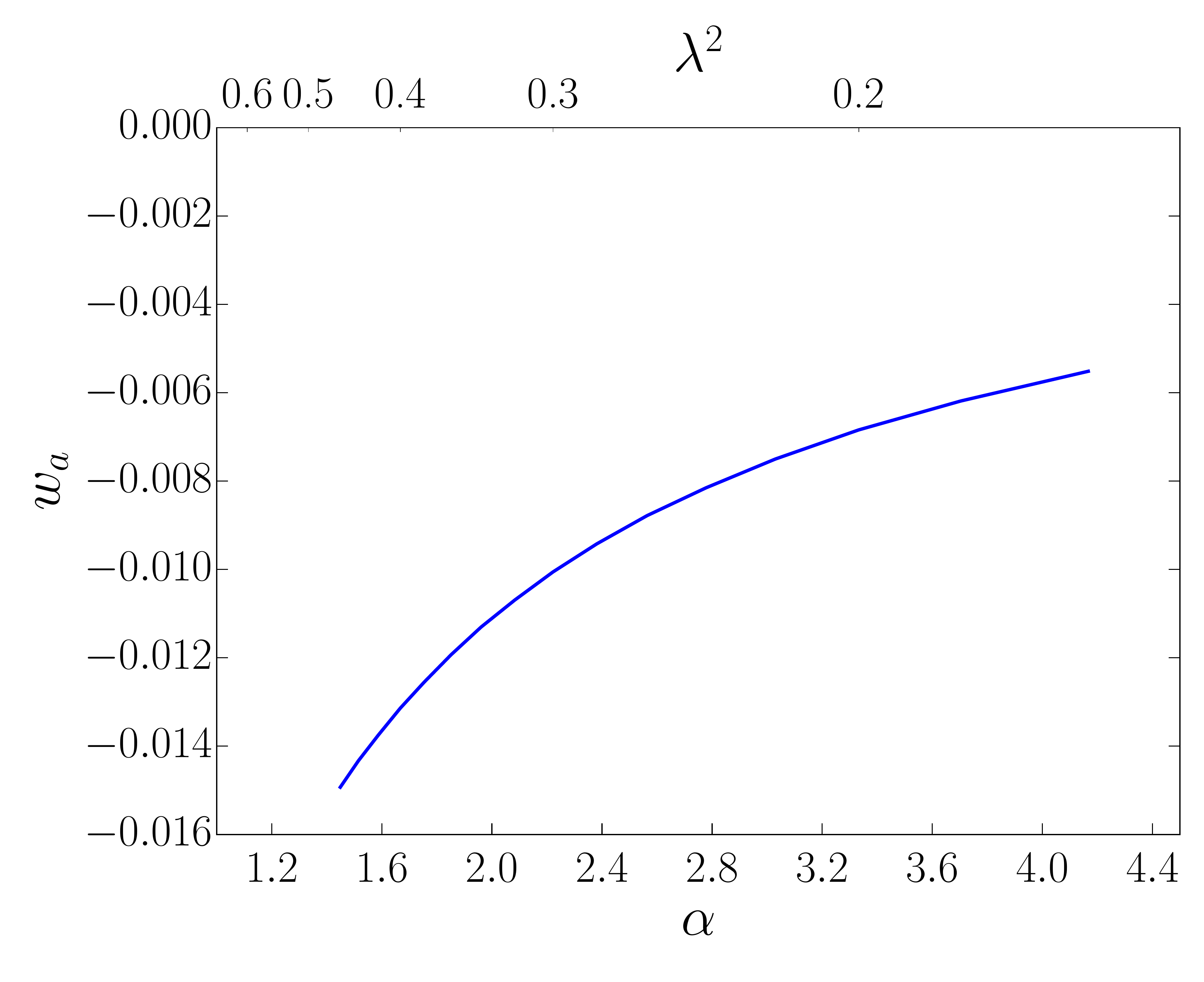}
	\caption{$w_a$ against $\alpha$ and $\lambda^2$ for \\
		\mbox{$\lambda^2<0.46\Leftrightarrow\alpha>1.45$}, values in the text are quoted to 2 s.f.}
	\label{fig:app_amendola2}
\end{figure}

\begin{figure}[]
	\centering
	\includegraphics[width=0.99\linewidth]{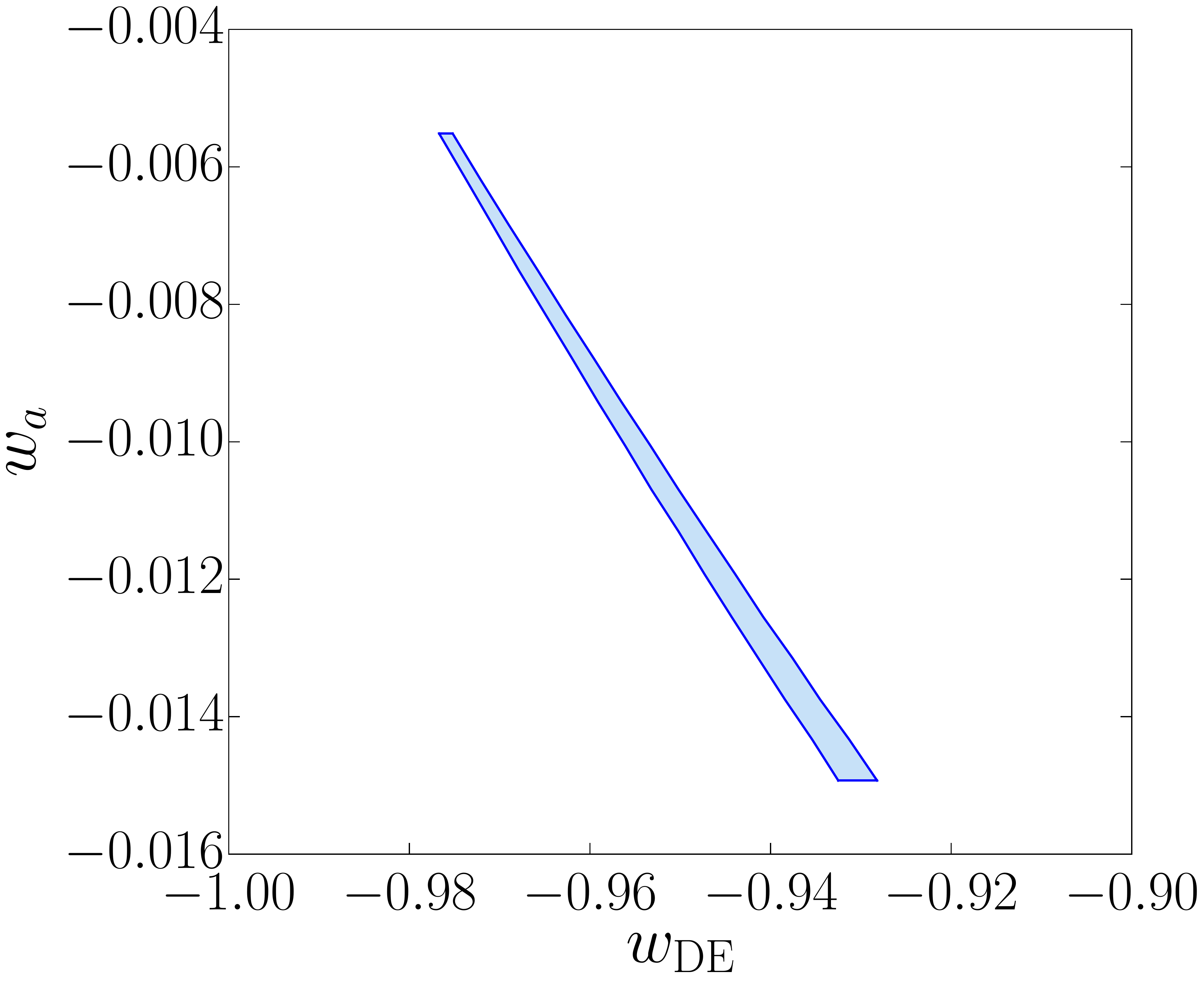}
	\caption{$w_a$ versus $w_{\mathrm{DE}}$. The allowed parameter space depicted lies well within the 1-$\sigma$ Planck contour.}
	\label{fig:app_amendola3}
\end{figure}

Using our Taylor expansion of $w_{\varphi}$ to first order, 
(cf.~Eq.~\eqref{eq:app_w_taylor}), we obtain a range of values for $|w_{a}|$ 
that are 
of $\mathcal{O}(10^{-2}) - \mathcal{O}(10^{-3})$. These values easily lie within 
current Planck bounds \cite{Ade2016}, but can be potentially observable in the near future, e.g. by EUCLID. 
This is illustrated in Figs.~\ref{fig:app_amendola2} and~\ref{fig:app_amendola3}.

The above are valid in general for exponential quintessnce. We now apply our 
findings to our quintessential inflation model with $\alpha$-attractors
We convert from $\lambda$ to $\alpha$, using 
\mbox{$\alpha=2/3\lambda^2$} (cf. Eq.~\eqref{eq:V_quint}), and 
restate all our findings in terms of $\alpha$.  We find that only values of 
$\alpha \geq 1.5$ accord with all the required Planck constraints and set an 
upper bound of $\alpha=4.2$ to avoid a super-Planckian $\phi$.
Figs.~\ref{fig:app_amendola2} and \ref{fig:app_amendola} are labelled both in 
terms of $\lambda^2$ and $\alpha$. 

%
\clearpage

\onecolumngrid

%

%

\end{document}